\documentclass{pasa}%

\title[Multiwavelength Observations of AB Doradus]{Multiwavelength Observations of AB Doradus}
\author[O.B. Slee et al.]{O.B. Slee$^1$, N. Erkan$^2$, M. Johnston-Hollitt$^3$ \and E. Budding$^{3,4}$
\\
\affil{$^1$Australia Telescope National Facility, CSIRO, Australia}%
\affil{$^2$Canakkale Onsekiz Mart University, Canakkale, TR 17100, Turkey}%
\affil{$^3$School of Chemical and Physical Sciences, Victoria University of Wellington, New Zealand}%
\affil{$^4$Carter Observatory, Wellington, New Zealand}}%
\jid{}
\doi{}
\jyear{}

\begin{document}%
\begin{abstract}
We have observed the bright, magnetically active multiple star AB Doradus in a multiwavelength campaign centring around two large facility allocations in  November  2006 and January, 2007.  Our observations have covered at least three large flares.  These flares were observed to produce significant hardening of the X-ray spectra during their very initial stages.  We monitored 
flare-related effects using the
Suzaku X-ray satellite and  the Australia Telescope Compact Array (ATCA) at 3.6 and 6~cm.  Observations at 11 and 21~cm were also included, but they were compromised by interference.  Optical monitoring was also provided by broadband $B$ and $V$ photometry and some high-dispersion spectrograms.  From this multiwavelength coverage we find that the observed flare effects can be mainly associated with a large active region near longitude zero.  The second major X-ray and microwave flare of Jan 8, 2007 was observed with a favourable geometry that allowed its initial high-energy impulsive phase to be observed
in the higher frequency range of Suzaku's XIS detectors.  
 
The fractional circular polarisation (Stokes V/I) was measured in the uv data for the complete runs, for 25 min integrations and, at 4.80 GHz, for 5 min integrations, using the radio data of Nov 21 2006 and Jan 08 2007.  Most of the full data sets showed V/I fractions from AB Dor B that were significant at greater than the 3$\sigma$ level. In several of the 5 min integrations  at 4.80 and 8.64 GHz this fraction reached a significance level between 3 and 9$\sigma$. Lack of angular resolution prevented identification of these high V/I values with one or other of the two low-mass red-dwarf components of AB Dor B.
\end{abstract}
\begin{keywords}
astrophysical data-- multiwavelength observations; stars-- cool, active;  stars-- individual: AB Dor
\end{keywords}
\maketitle%
\section{Introduction}
\subsection{Cool star activity}
Magnetic activity drives high-energy processes in the atmospheres of stars. Powerful releases of energy (flares) are regularly observed and display plasma temperatures of several 10$^7$ K, reaching up to $\sim 10^8$ K in extreme cases (see e.g., G\"{u}del et al., 1999; Maggio et al., 2000; Favata et al., 2000; Franciosini et al., 2001).  Such processes give rise to observable effects in the microwave region, and while the high-energy tail of thermal electrons has been observed in hard X-rays (Pallavicini \& Tagliaferri, 1999; Favata \& Schmitt, 1999; Franciosini et al., 2001), evidence on the high-energy non-thermal electron population has generally been scant. 

One of the important facts to emerge from soft X-ray (SXR) observations during the Skylab era was the recognition that magnetic loops constitute the basic `building blocks' of coronal structure (Rosner et al., 1978). Using
the solar analogy, it became clear by 1980 that loops play a basic role in coronal energetics in both active and quiescent phases  (e.g. Pallavicini et al., 1977, Kane et al., 1980).
Regarding flares, according to the evaporation model (cf.\ Antonucci et al., 1984), beams of fast electrons travel along loops and deposit their energy towards the lower chromospheric material. Heated plasma may subsequently expand into the loop structures. This hot plasma cools through conduction and radiation, but more than one emission or absorption mechanism is likely to be involved (Dulk et al., 1986, Bastian \& Gary, 1992, Alissandrakis et al., 1993).
In the cm range, however, the dominant emission usually involves the gyrosynchrotron process from loops.

The solar model is often used in trying to understand the mechanism and location of flares in active stars.   A non-thermal electron population is thought to be present in the impulsive phase of flares (Doschek, 1972) and this is relevant to the power-law index of the electron energy distribution and its lower energy bound (Phillips \& Neupert, 1973), used in microwave data analysis.   Hard X-rays ($>$10~keV) should be emitted during the initial stages of the flare.  Softer X-rays become detected as high energy electrons impinge on denser (`chromospheric') layers as the flare develops, giving rise to thermal bremsstrahlung. Complementary microwave observations should evidence the non-thermal electron source, giving information on its energy spectrum and possible changes during the course of a flare. In fact, radio data have sometimes pointed to a very strong non-thermal electron component (e.g., Lim et al., 1994). 

  The present study concentrates much of its in-depth analysis
on such microwave data from the active star AB Doradus  in an effort to gain clearer insights into 
how  such large energy releases occur.
 Within the two days of combined X-ray and radio data we discuss, the latter showed higher signal to noise (S/N) ratio for typical integration intervals of several minutes, particularly in the higher frequency bands.

The multi-wavelength campaign that stimulated the present work was originally aimed at examining,  in particular, the initial population of highly accelerated (non-thermal) electrons. While very hard, non-thermal X-rays would be too weak to detect in quiescence, they were calculated to be picked up by Suzaku's sensitive detectors during the onset of large flares. Simultaneous radio and X-ray observations then have the possibility to reveal the `Neupert effect' in flares.  This concerns the correlation between radio and soft X-ray or UV flare emissions (Neupert, 1968), in which the short wavelength light curve of radiatively cooling plasma tends to follow the integral of the radio emission.   This can be interpreted in terms of `loss cone' gyrosynchrotron-emitting electrons encountering denser chromospheric layers, whereupon X-ray emission follows the radio decline with a short time lag.  The Neupert effect has been observed in several stars using radio and optical U-band data (Hawley et al., 1996; G\"{u}del et al., 1996; 2002a,b), and also radio and X-ray data (Osten et al., 2004).     

This multiwavelength surveillance has allowed checking  of rotational modulation effects.  In particular, AB Dor is associated with long-lived centres of surface activity, noticed by persistent broadband (optical) maculation.  The relationship of electron energization to such activity centres can be explored.  In this way, optical photometry and high-resolution spectroscopy, observed over intervals around the time of the X-ray and radio observations, supplement the overall picture of this intriguing star.

\subsection{AB Doradus} 

AB~Dor (HD~36705, K1V-type, $V$ = 6.93 mag, $M$ = 0.76 $M_{\odot}$) is a well-studied magnetically active young star that has been observed over a wide range of wavelengths. The main target is the primary component of a small multiple system containing at least 4 stars.  Historically, AB Dor A was regarded as the primary of the visual binary Rst 137 (separation around 7.5 arcsec).  More recently both AB Dor A and B were shown to be both binaries, involving low-mass companions (see e.g.\ Janson et al., 2007, for a review giving modern parametrization).  

The K-type main star, at 14.9~pc distance (Perryman et al., 1997), displays numerous strong flares on time scales from minutes to weeks, with two extreme cases having integrated X-ray fluxes of $\sim$4$\times$10$^{-9}$erg cm$^{-2}$ s$^{-1}$, i.e.\ about 60 times the quiescent level. Observed flaring occurs at a mean rate of about one per day. Peak temperatures of around 110$\times$10$^6$~K (Figure~1 in Maggio et al., 2000) were inferred. This main star, AB~Dor Aa, has the relatively fast mean rotation of 1.40$\times$10$^{-4}$ radians~sec$^{-1}$ ($\sim$12.3~h period) and was long associated with surface `spots' that allow not only the mean rotation, but also differential rotation and cyclic behaviour to be examined 
(J\"{a}rvinen et al., 2005).

The conclusion of Rucinski (1983) and Innis et al.\ (1986) that AB Dor Aa (subsequently, usually just `AB Dor') may be a pre-main sequence star was questioned by Micela et al.\ (1997).  Still, AB Dor turns out to be very probably  near to the zero-age main sequence (ZAMS).  Zuckerman et al.'s (2004) checking of the galactic space motions of the Moving Group that contains AB Dor indicates an age of about 50 My. The youth of the system was confirmed in part by lithium abundance studies (Rucinski, 1982, Hussain et al., 1997).
 
There are various markers of AB Dor's magnetic activity.   As well as the short-term `starspot' behaviour, photometric studies over long periods have shown the average brightness goes through cycles over the years. Observations over the period 1978-88, showed the star's average brightness to decrease steadily, reaching a minimum by the end of this interval.  Subsequently, increases over the years 2000-2004 were observed and a maximum level reached (Amado et al., 2001; J\"{a}rvinen et al., 2005).  This longer-term variation has been seen as an indicator of solar-type cycles.
Another indicator of AB Dor's magnetic fields is its active chromosphere.  This is marked by the presence of strong CaII and H alpha emission (Bidelman \& MacConnell, 1973; Vilhu et al., 1987; Budding et al., 2009).  As well, the active corona gives rise to strong and variable radio and X-ray radiation (Pakull, 1981; Slee et al., 1986),  although 
the evidence from different spectral regions is not
clear-cut regarding indications of activity (K\"{u}rster et al., 1997).

  The presence of enlarged coronal structures on AB Dor has been discussed at least since the
giant loop model of Mestel and Spruit (1987), as well as the observational confirmation of emitting material linked to plasma at heights of order
several stellar radii above the photosphere by Collier Cameron and Robinson (1989).  Loop sizes
were inferred from analysis of the fading of X-ray flares observed with the BEPPOSAX satellite by Maggio et al.\ (2000), but flaring-loop heights were, in that case, taken to be somewhat smaller ($H\lesssim$  $0.3 R_{\rm star}$), since the flares were observed to continue through all the rotation cycle without showing any eclipse effects, so that their source should not be far above the polar regions. The rotation axis is believed to be inclined at $\sim$60$^{\circ}$ to the line of sight.  Other methods have tended to
support this scale of size for X-ray emitting structures
(Sanz-Forcada et al., 2003; Hussain et al., 2005.) 

 A  comprehensive assessment of coronal energization, involving X-ray observations repeatedly collected over a three year period from the XMM-Newton and Chandra satellites was given by Sanz-Forcada et al.\ (2003)
 and summarized by Sanz-Forcada et al.\ (2004).
  Loop sizes were inferred to come in a distribution of sizes, but mostly relatively small compared to the stellar radius.  The high energy tail of electron energy distributions 
was found to vary significantly with the general activity state of the star.  Hussain et al.\ (2007) combined evidence from the rotational modulation of the X-ray flux
with high dispersion ground spectroscopy to 
produce detailed maps of likely emission configurations.
Possible geometries of coronal emission regions were also considered by Lim et al.\ (1994), who
examined radio data-sets comparable to ours.
Evidence on flare region geometry is, as a whole, rather incomplete and often somewhat indirect, however. 
Modelling tends to involve many adjustable parameters,
implying over-determined curve-fittings and ambiguous results. 

 AB Dor has been the subject of a number of previous
 satellite-based studies: for example, Rucinski et al.\ (1995); Mewe et al.\ (1996); K\"{u}rster et al.\ (1997); 
 Vilhu et al.\ (1998);  Schmitt et al.\ (1998); 
Maggio et al.\ (2000) and others, as discussed by
Sanz-Forcada et al.\ (2003).
Repetitive phase-linked phenomena have demonstrated
the presence of localized inhomogeneities,
with effects involving large-scale photospheric magnetic field concentrations,
through to chromospheric and transition region condensations and on to hot coronal plasma volumes
 typically in the range of  several to a few tens of millions of K -- occasionally more extreme.
 Recently, Lalitha \& Schmitt (2013) have considered the
 relationship between trends of X-ray emission and 
 optical mean flux, associated, tentatively, with a 17 y period.
 Lalitha \& Schmitt found that the high basal level of X-ray emission associated with the strong activity of AB Dor A
 does vary in response to the expected long-term cycling,
 but the amplitude of this background emission is not so great
 compared with other known cases, in particular the Sun.
 
So although the solar example continues to underlie many of the concepts, as with quiescent and active conditions,
the situation of AB Dor can involve structures and energies at least a
couple of orders of magnitude greater.  Alternative scenarios can be applied to the emission mechanisms, and further clarification of this subject forms a strong motivation for the present work.

The cyclic, quasi-repetitive character of AB Dor's radio behaviour, over long periods of time, confirms the existence of structures maintaining the same longitude.  The sources are explained in terms of field concentrations that slowly migrate over a number of years.  Lim et al.\ (1994), from studying radio emission peaks and simultaneous optical data, deduced the emission to arise from close to the spotted areas.  Many repeated phenomena from photosphere, chromosphere and corona thus point to the presence of `active longitudes'.
Notwithstanding the foregoing argument on flaring loop heights, there is also evidence of very extended coronal condensations.  For example, the FUSE (Far Ultraviolet Spectroscopic Explorer) and HST (Hubble Space Telescope) have observed transition region lines (CIV 1548\AA, SiIV 1393\AA\ and OVI 1032\AA, $T \sim10^5$ K), with speeds of up to 270 km  s$^{-1}$.  An extended region of optically thin plasma, with high-level co-rotation, at heights  up to 2.6 $R_{\rm star}$ was then suggested (Brandt et al., 2001; Ake et al., 2000).

Doppler Imaging techniques were often applied to AB Dor (K\"{u}rster et al., 1994; Unruh and Collier Cameron, 1994; Donati and Collier Cameron, 1997; Donati et al., 1999, Hussain et al., 2002). These studies of the surface distribution of active regions have demonstrated the persistence of two dominant spot longitudes, but at different latitudes.  Donati and Collier Cameron's (1997) account of AB Dor from optical cross-correlation techniques concluded that the scale of differential rotation was approximately 40 times greater than that of the Sun.  This would then suggest a preferred latitude for spots (in a given hemisphere), or the differential rotation would soon produce different phases for the recurrent maculation wave on which the ephemeris and consequent longitude system is based. 

\section{Observations}

\subsection{X-ray observations}

Launched in 2005, Suzaku is engaged in observations of astronomical X-ray sources that form  part of a joint programme of the Japanese Institute of Space and Astronautical Science (ISAS-JAXA) in collaboration with NASA-GSFC and other organizations (cf.\ Mitsuda et al., 2007).  Three of Suzaku's four X-Ray detectors of the imaging spectrometer (XIS) were operable during our allocations.  A hard X-ray detector (HXD) sensitive over the energy range 0.2-600 keV  with low background and an angular resolution that should allow application to research on brighter flare stars
was also available.
A Suzaku proposal for observations of AB Dor was thus made in early 2006 with an aim to research
the high-energy non-thermal electron population, particularly during the initial flare impulse.  Possible detection of very hard, non-thermal X-rays could allow better understanding of the triggering mechanism.  

The X-ray observations, made simultaneously with the radio ones, were carried out in November 2006 and January 2007. 
Unfortunately, the signal from the HXD during these observations did not permit reliable analysis (Arnaud, private communication). However, the XIS (intermediate energy) data show interesting effects.  

The XIS has a field of approximately 18 arcmin square covered by the three available detectors XIS0, 1, and 3. 
  The angular resolutions of the Suzaku telescopes range from 1.8 to 2.3 arcmin half-power diameter. This resolution does not significantly depend on the energy of the incident X-rays in the available range, but it entails that all four known components of the
 multiple star are covered by the acceptance beam.
The archived Chandra data from 2004  (http://cda.harvard.edu/ pop/),  has much higher angular resolution
and it shows two separate sources with both a 52 ks exposure using the HETG, and an 85 ks exposure with the LETG grating, in the positions of AB Dor A and B on the sky.  
Aperture photometry of these images shows the AB Dor A source 
(as enclosed within a 1.7 arcsec radius aperture)
to have an average flux density of 83.3\% of the sum of the two sources with the HETG instrument, and 87.8\% with the LETG.
Since these exposures cover time intervals of the order of a day, there is a good chance that they
would have included a flare, but mostly the output would be dominated by accumulated quiescent 
emission from the two heated coronae in the proportions 5.7 to 1, for AB Dor A to B.  
 
 The simultaneous, higher angular resolution, radio observations show that three large flares occurred from AB Dor A during the 
 joint coverage. The X-ray cameras registered distinct 
 flares  around the same times. Further insights are gained when the X-ray and microwave observations are superposed, as discussed in Section 4.
 
\begin{figure*}
\begin{center}$
\begin{array}{c}
\hspace{0cm}
\includegraphics[scale=1.1, angle=0]{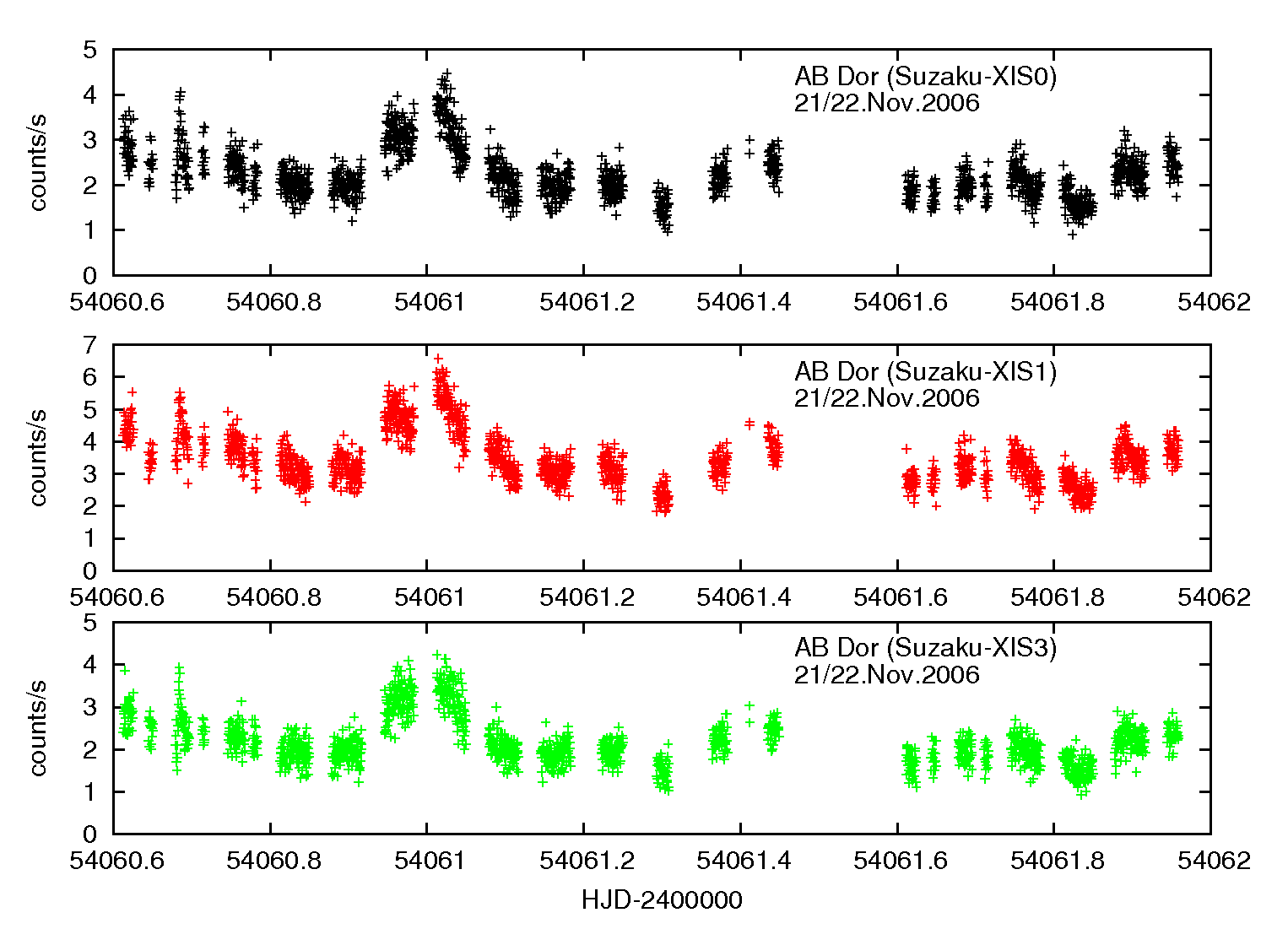} \\
\hspace{0cm}
\includegraphics[scale=1.1, angle=0]{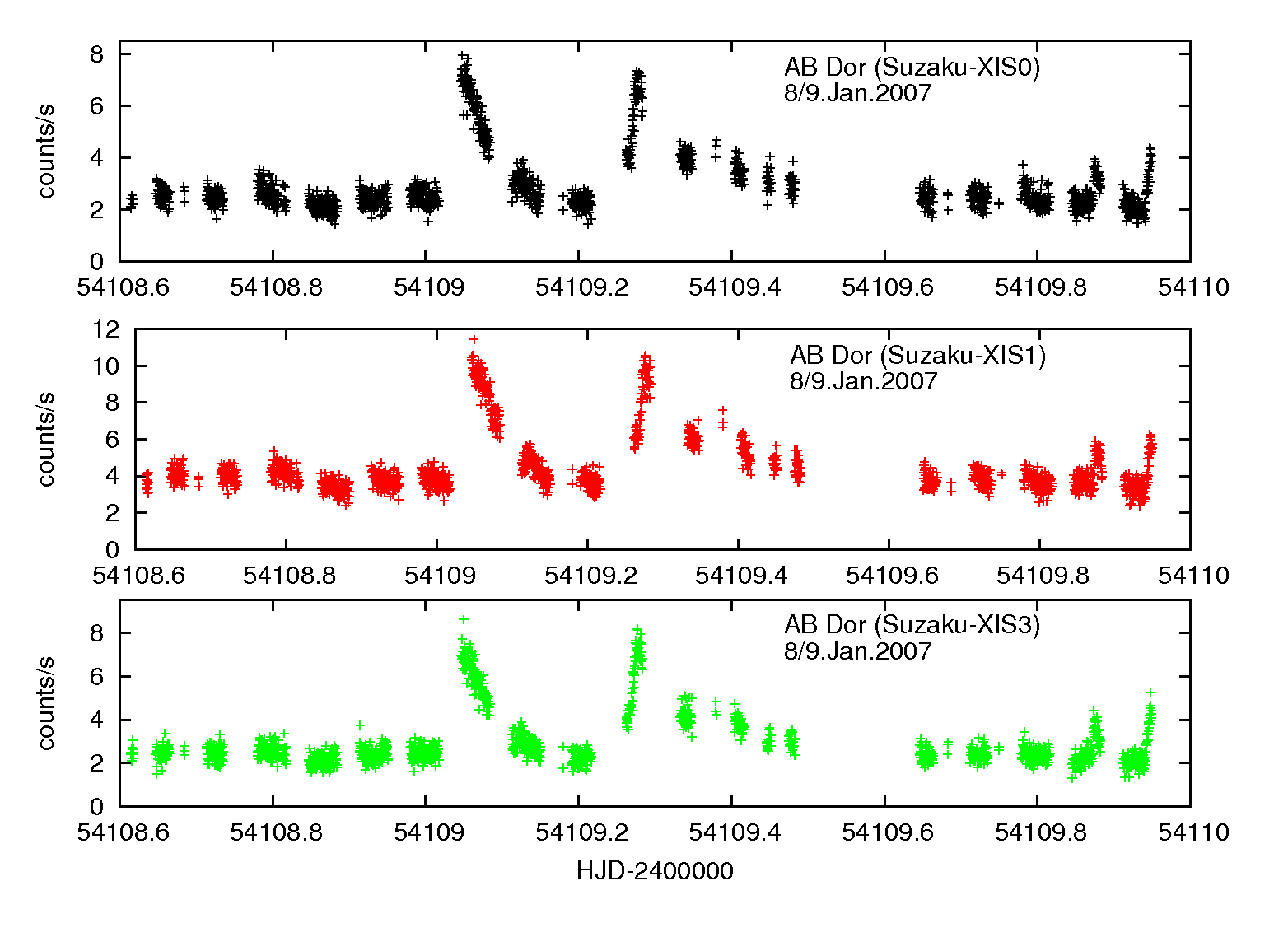}
\end{array}$
\caption{Light curves of AB Dor collected by the Suzaku XIS detectors; top:  for Nov 21-22, 2006; and bottom for Jan 8, 2006.  These X-ray raw data from the intermediate range cameras integrate photon energies between 0.3 to 10 keV. The $y$-axis count units are linearly proportional to the received flux.}
\label{AB_Dor_f1}
\end{center}
\end{figure*} 

The X-ray data were taken from the HEASARC archive (NASA,  cf.\ White, 1992).  HEASoft reductional software, including {\sc FTOOLS} and {\sc XANADU}, is similarly available.  In Figure~1  we show the integrated XIS raw data  over the range 0.3 to 10 keV from the two observation dates.
The original data were collected on 2006 Nov 21-22 (HJD 2454060-1; 2h 43m 51.9s until 11h 01m 35.2s, i.e.\ $\sim$116000 s) and on 2007 Jan 8-9 (HJD 2454108-9; 2h 46m 25.2s until 10h 48m 3.7s, i.e.\ $\sim$115000 s).

\begin{figure*}
\vspace{0cm}
\begin{center}
\hspace{-1cm}
\includegraphics[scale=0.27, angle=90]{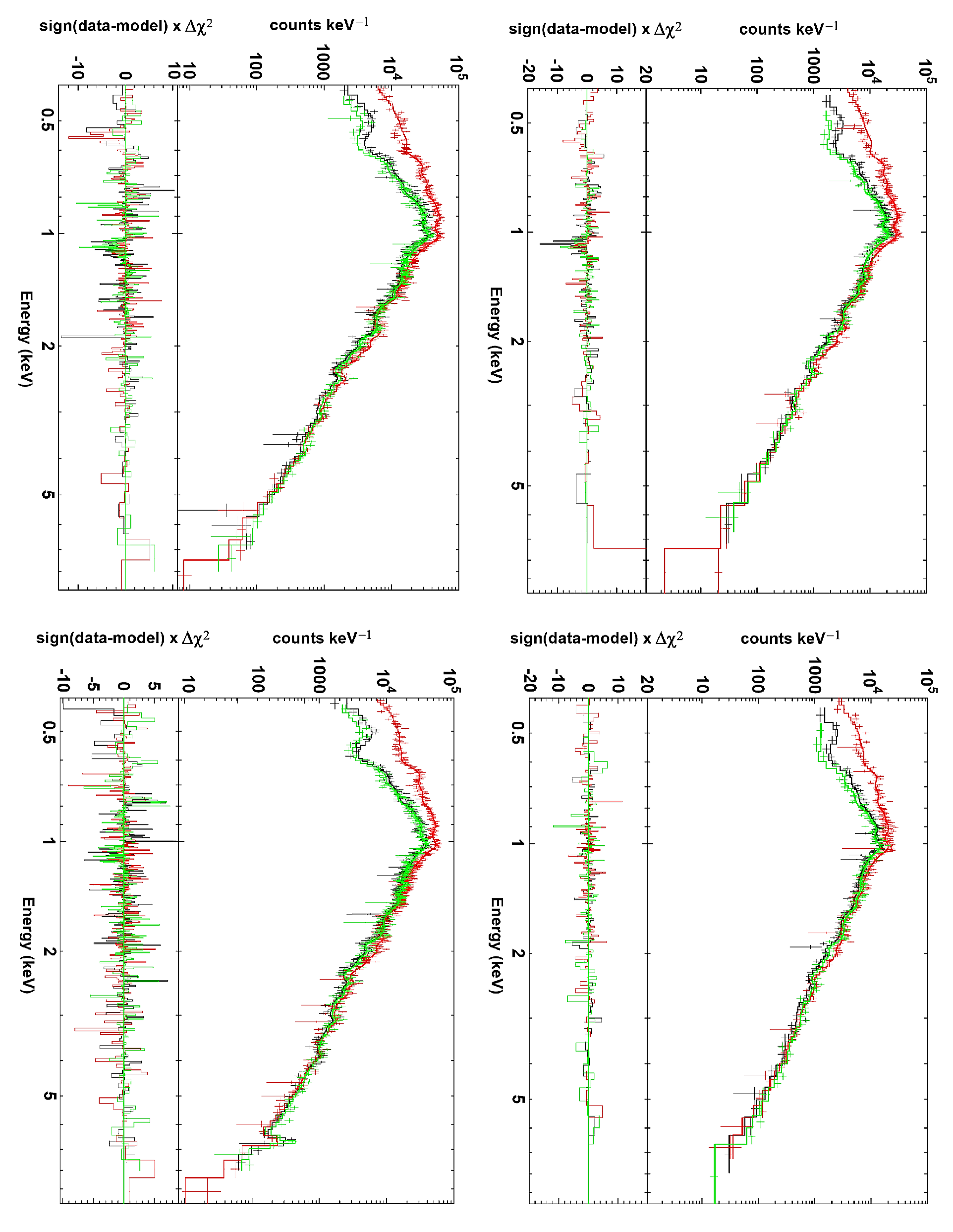} 
\caption{{(top) AB Dor, 21/22 November 22, 2006, Suzaku X-ray spectra derived from integrations with the XIS1 (upper trend of points), XIS0 and XIS3 (middle and lower) cameras, respectively. The left panel corresponds to the quiescent flux,
the right to flaring. (bottom) The same arrangement but for 8/9 January, 2007. The Fe XXV emission at around 6.7 keV  feature is only apparent at times during flares.}}
\label{AB_Dor_njspec}
\end{center}
\end{figure*}

 Individual 30 s integrations with the separate cameras during quiescent intervals show a typical scatter of about 12\%\ of measured levels. 
This implies that time averaged values covering a few hours of quiescence would reduce the error of a representative mean to a few percent.  We can 
check this in the foregoing averaged values, i.e.\
$f_{\rm av} = 4.71 \pm 0.04 \times 10^{-11}$ erg cm$^{-2}$ s$^{-1}$, for the  Nov data and 
$f_{\rm av} = 6.52 \pm 0.13 \times 10^{-11}$ erg cm$^{-2}$ s$^{-1}$, for the more active Jan data.
The X-ray spectra matching these net fluxes, with their
modellings, are presented in Figure~2 (left). 
  
\begin{figure*}
\begin{center} $
\begin{array} {c}
\hspace{-0.8cm}
\includegraphics[scale=1.2, angle=0]{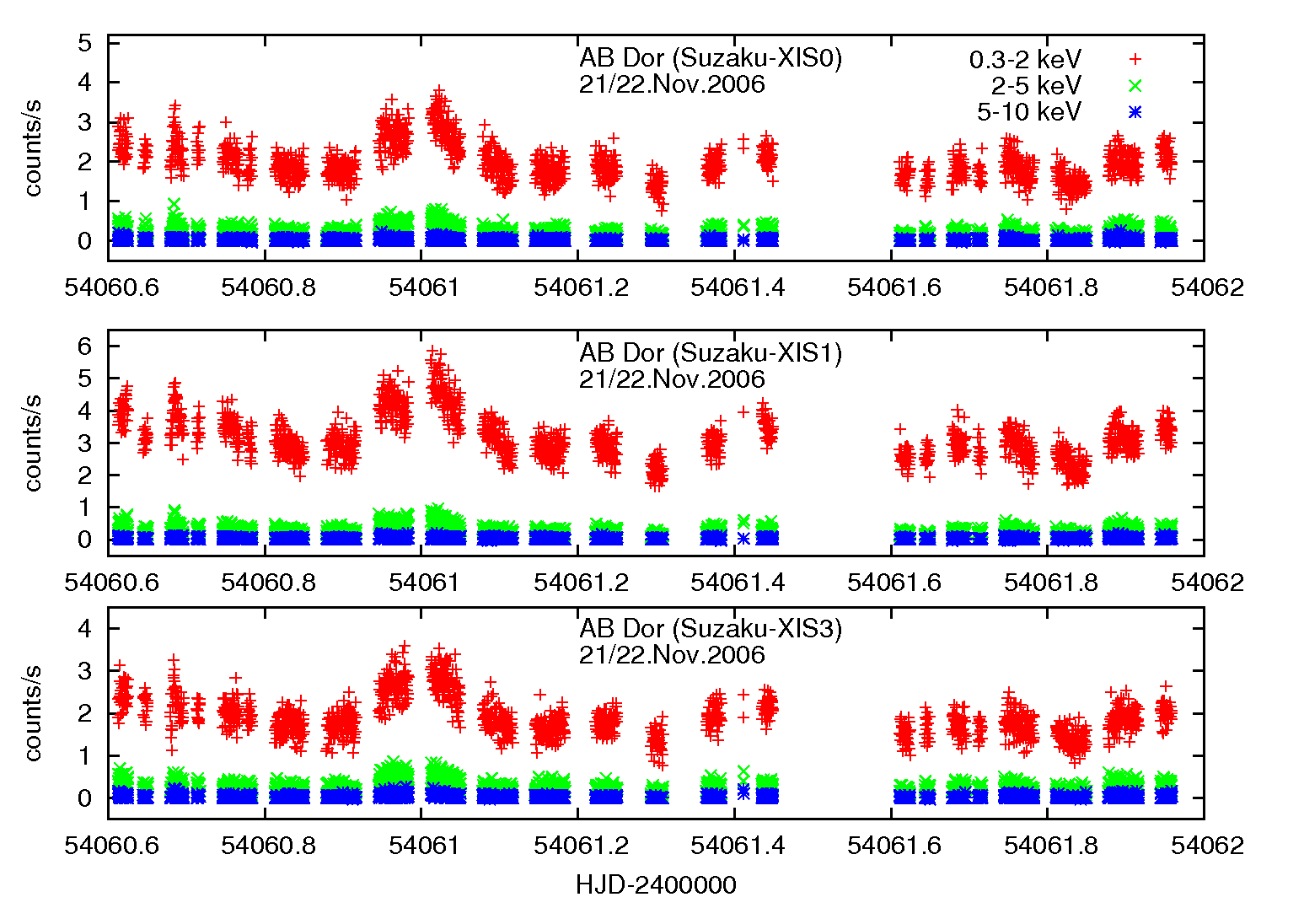} \\
\hspace{-0.8cm}
\includegraphics[scale=1.2, angle=0]{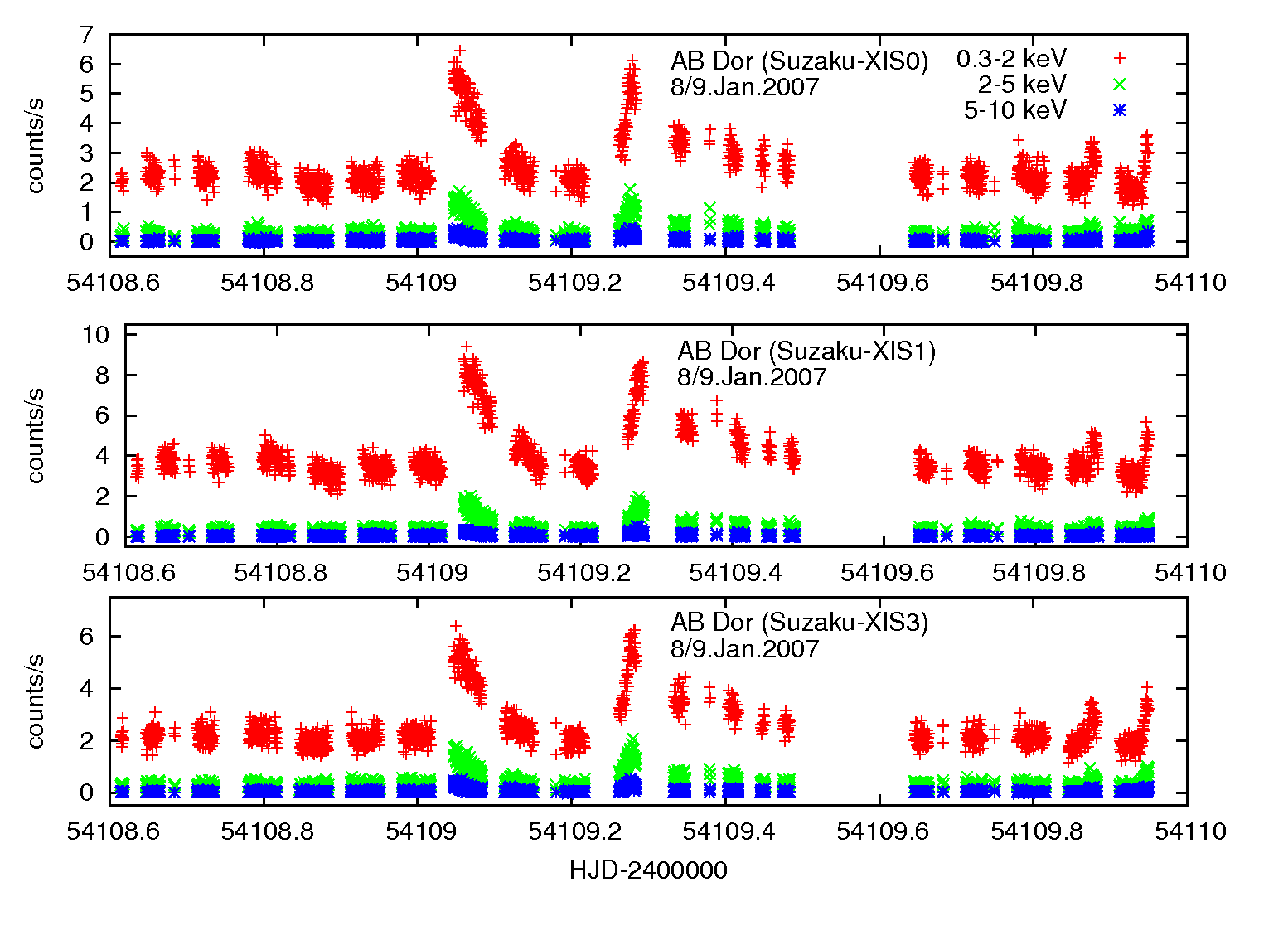}
\end{array}$
\caption{(top)AB Dor, 21/22 November, 2006, Suzaku X-ray observations for different energy ranges   
(see Section 2.1 text). Data from the XIS1 camera (lowest energy range) are shown as the upper trend of points (red), with middle (green) and lower (blue) trends corresponding to the (higher energy) XIS3 and  XIS0 cameras respectively (bottom).   Related TBabs and modelling parameters are referred to in the text.  The same arrangement but for 8/9 January, 2007.}
\label{AB_Dor_xfig5}
\end{center}
\end{figure*}

\begin{figure*}
\begin{center} $
\begin{array} {c}
\hspace{-0.8cm}
\includegraphics[scale=0.45, angle=-90]{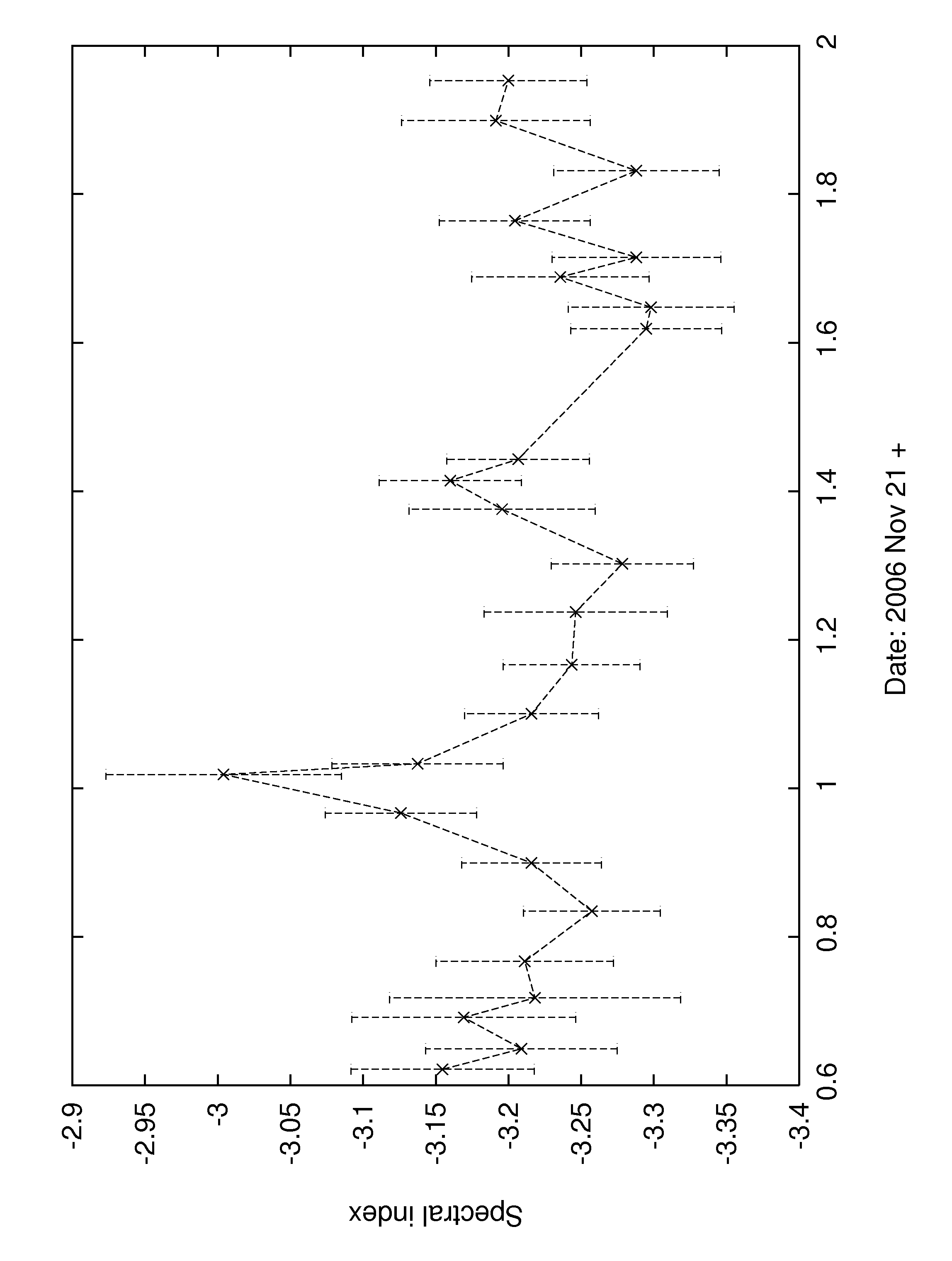} \\
\hspace{-0.8cm}
\includegraphics[scale=0.45, angle=-90]{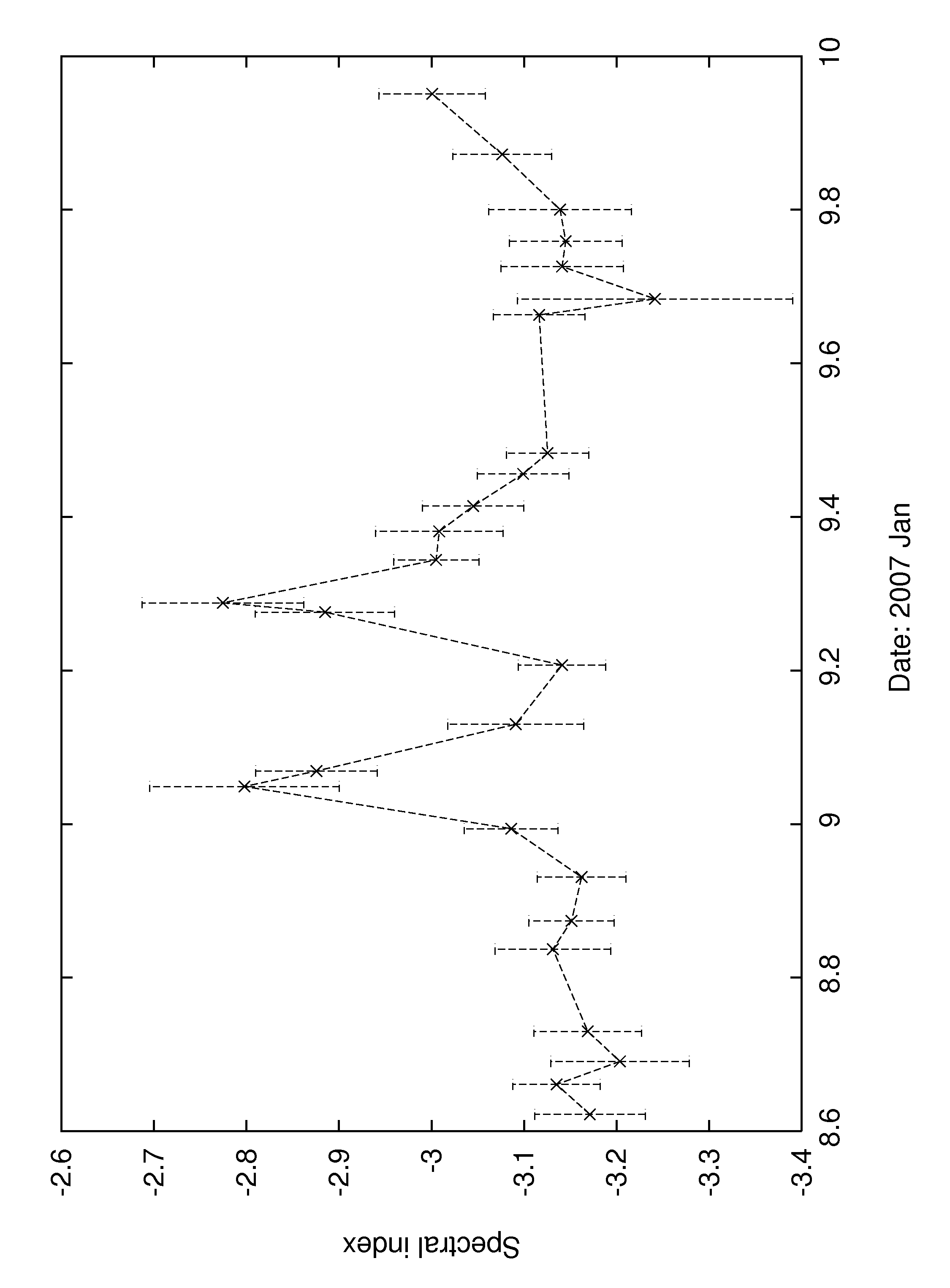}
\end{array}$
 \caption{AB Dor, 21/22 November, 2006, spectral index variation corresponding to Figure~3 (top) and for 8/9 Jan, 2007 (bottom).}  
\label{AB_Dor_xfig5a}
\end{center}
\end{figure*}

The XIS spectra were processed using {\sc XANADU}, particularly its {\sc XSPEC} spectral-fit program.
This calls on the Tuebingen-Boulder ISM absorption model {\sc TBabs}, which accounts for the intervening effects of the interstellar medium.  Since the modelling is overdetermined by the available parameter-set, our approach was minimalistic
in trying to find the smallest number of independent parameters
consistent with the data. The processing involves use of the {\sc MEKAL} (MEwe-KAastra-Liedahl) code, constructing multi-temperature plasma emission models (Kaastra et al., 1996). This  analysis was carried out by one of us (NE) whilst on study leave at the Armagh Observatory. 

 X-ray fluxes from stellar coronae are characterized by a distribution of emission measures (EMD)
for the various contributing species to the net emission.
The flux normally shows a distinct maximum, with a 
corresponding brightness temperature $T_{\rm max}$,
typically in the order of 10 MK.  
The brightness temperature $T$ can be related to
the EMD through a proportionality of the form ($T$/$T_{\rm max})^{\alpha}$.  {\sc MEKAL} model parameters include $\alpha$ and $T_{\rm max}$, $nH$,  abundances of the elements, and a normalization parameter.  In the present cases we derived (from optimal curve-fitting) $T_{\rm max} = 2.588 \pm 0.039$ (keV) and   $T_{\rm max} = 3.129 \pm 0.046$ (keV, with s.d.\ error measures here and elsewhere), i.e.\ about 9.3 and 11.2$\times$10$^6$ K, for the November and January allocations respectively.  Acceptable fits were
found with  $\alpha$ set to 0.1, the corresponding low variation of the EMD with brightness temperature,
indicative of dense and compact conditions
for the spectral region observed.
  Further details on the reduction and modelling procedures were given by Erkan (2012). 

The spectral analyses covered all three cameras' data and the results were cross-checked for self-consistency. For the range 0.3 to 10 keV energy,
{\sc TBabs}  modelling of the quiescent levels gave the following flux 
averages:


\vspace{2ex}

\begin{tabular}{c}
21/22 November, 2006 \\
$f_{\rm XIS0} = 4.69 \times 10^{-11}$ erg cm$^{-2}$ s$^{-1}$;\\
$f_{\rm XIS1} = 4.77 \times 10^{-11}$ erg cm$^{-2}$ s$^{-1}$;\\
$f_{\rm XIS3} = 4.65 \times 10^{-11}$ erg cm$^{-2}$ s$^{-1}$;
\end{tabular}

\begin{tabular}{c}
8/9 January, 2007 \\
$f_{\rm XIS0} = 6.34 \times 10^{-11}$ erg cm$^{-2}$ s$^{-1}$;\\
$f_{\rm XIS1} = 6.55 \times 10^{-11}$ erg cm$^{-2}$ s$^{-1}$;\\
$f_{\rm XIS3} = 6.70 \times 10^{-11}$ erg cm$^{-2}$ s$^{-1}$;
\end{tabular}
\vspace{2ex}

The high energy range tail of the X-ray spectrum 
is characterized 
by an index $\delta$, where $F_{\nu} \propto \nu^{\delta}$ (Dulk, 1985) and $\delta \approx -3$ is often observed in practice\footnote{Dulk (1985) actually wrote $F_{\nu} \propto \nu^{-\delta}$.}. Our results are similar to those found for other active cool stars 
(e.g.\ Osten et al., 2002), showing a peak at about 1 keV and a high energy tail with a power-law slope close to --3 in the range 1-5 keV.  Energy resolution beyond 5keV is less distinct, although the 6.7 keV Fe emission feature becomes clear during flares: at other times it was not detected, as shown in Figure~2.

Data analysis  involves the so-called Good Time Intervals (GTI), selected
according to investigator preference
but generally in accordance with suitable signal levels. These intervals may be extended over times of lesser scatter, 
or perhaps flaring episodes.
Figure~3 presents reduced observations of AB Dor. GTIs  are here ascribed simply 
in accordance with the more continuous on-time integrations of the source on the two observation dates. 
 Flux-weighted averages for the three bandpasses (low, medium
 and high, say)
 in Figure~3 correspond to 2.42, 6.91 and 14.7$\times$10$^{16}$ Hz as their effective frequencies. The low and medium energy ranges allowed clear measures of the X-ray incident flux throughout the monitoring.  Spectral indices can then be derived and 
compared with the average spectra shown in Figure~2. 
Data from GTI regions 8-10 in Nov 06 and 9-17 in Jan 07 correspond to times when prominent flares were observed. 

 Flux levels in the highest frequency channel (5-10 keV) had
very low signal strength throughout the monitoring, with
S/N ratios for individual integrations of typically $\sim$1.5, during quiescence, rising to about 3 near the flare peaks.
 These low levels of the high energy signal are consistent,
 when integrated over the selected time intervals,
  with the power-law form of the high energy tails in Figure~2. 
  In the flare regions, the highest energy channel shows distinctly more noticeable flux in Figure 3.  These flux increases across the medium-high range 
 are indicative of spectral hardening.  This can be seen in Fig 3 with GTIs 8 and 9 on Nov 06, but more so near the flare peaks in GTIs 9 and 12 of Jan 07, particularly the latter.
  
  Since the flux distributions on either side of the maximum
 tend to a power-law-like form (albeit with 
  different indices)
 flux-versus-frequency integrals tend to be dominated
by one of the limiting ordinate values. 
This means that the low energy range fluxes
can be scaled by a suitable constant to support estimates of the spectral power law index for the high energy tail. 
This was checked
empirically, and the scaling factor 3.2 was found to 
allow the mean value of $\delta$ from the low to medium flux ratio to agree with that for the medium to high range. In this way, the standard deviation of the $\delta$ values 
were generally reduced from about 0.25 to 0.08.  The corresponding
$\delta$ values, with their s.d.\ errors, are shown
in Figure~4. The sought differential variation of $\delta$ thus reflects the clear relative intensification of the 
medium flux measures during flares.  
  
    $\delta$ can be seen to approach --2.7 at the peaks in Fig~4. Noticeable also about the second flare of the Jan data is that this hardening occurs at the initial X-ray peak, but during the subsequent microwave peak and second slower enhancement the X-ray emission is softer.
 This corresponds to the classic Neupert effect, and the pattern is consistent with an initial high-energy triggering followed by a gyrosynchrotron flare with a later thermal cooling associated with electron precipitation.   
This point has interesting consequences for our appraisal of flare behaviour, and we return to it in the later discussion. 
 
\subsection{Radio observations}

\subsubsection{ATCA Data}

 Preliminary forms of AB Dor's radio light curves were given by Budding et al.\ (2009), mainly with the aim of showing that the various activity manifestations at different wavelengths were probably linked to identifiable centres of activity on the primary star.   
However, that was without detailed backgrounding of  the observational circumstances and data analysis that we now present in this and following
sections of the present paper.  In fact, the idea of repetitive phase-related enhancements in AB Dor's radio emission is not new. It was emphasized in the the study of Lim et al.\ (1994), where radio light curves showing a two-peaked structure that aligned with optically identified photospheric centres of activity was evidenced.  Lim et al.'s observations appear to have found the star in a more active condition than the present authors, however.  The peak intensities were significantly higher (by factors of typically $\sim$5) than those we discuss in what follows.  The emission turnover frequencies were also generally lower: our 8.6~GHz peaks being noticeably lower than those at 4.8~GHz. At the time of these observations the Australia Telescope Compact Array (ATCA) was an East-West synthesis telescope of total length 6~km operating through cm to mm wavelengths.\footnote{The facility was significantly upgraded in 2010 with the addition of the Compact Array Broadband Backend 
(CABB), enabling a 2 GHz observing bandwidth.} In the cm bands the array was then capable of simultaneous dual-frequency observations. In this experiment we operated at 1.384/2.368~GHz and 4.80/8.64~GHz in an attempt to explore the physical conditions responsible for the star's gyrosynchrotron emission over a large range of heights in its corona.

\begin{figure*}
\vspace{0cm}
\begin{center} $
\begin{array} {cc}
\hspace{-1cm}
\includegraphics[scale=0.40,angle=-90]{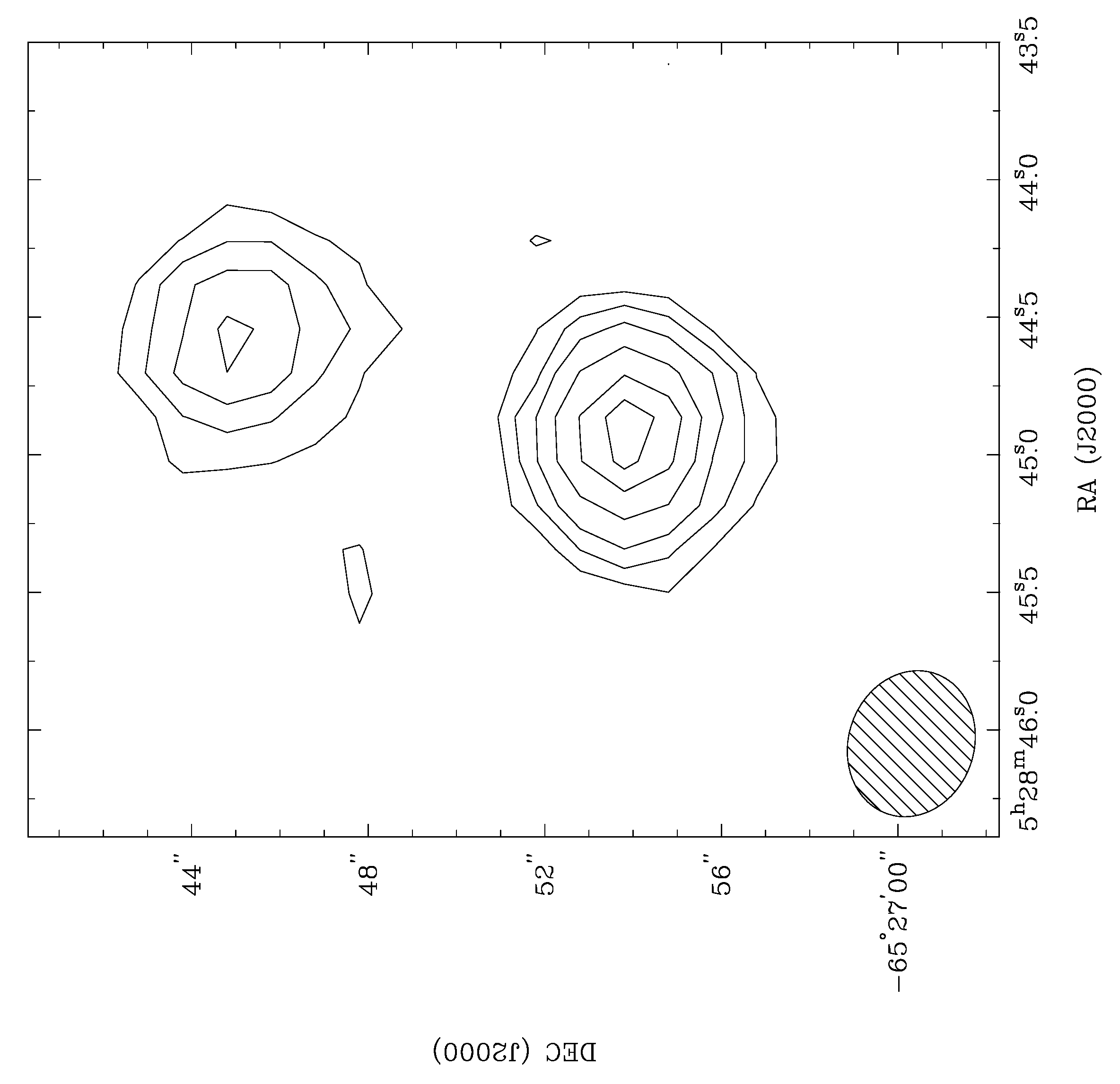} &
\includegraphics[scale=0.40,angle=-90]{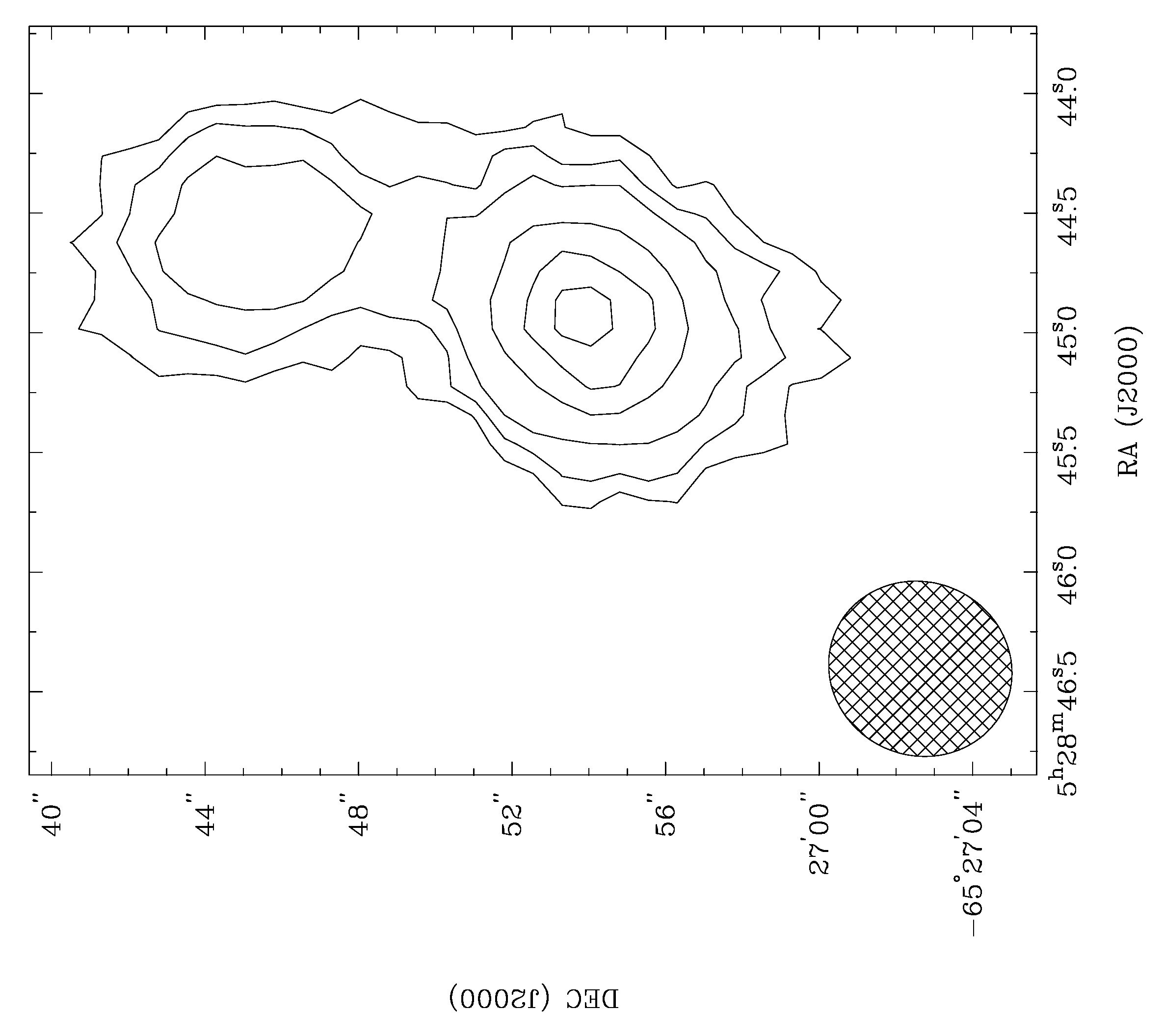} \\
\hspace{-1cm}
\includegraphics[scale=0.40,angle=-90]{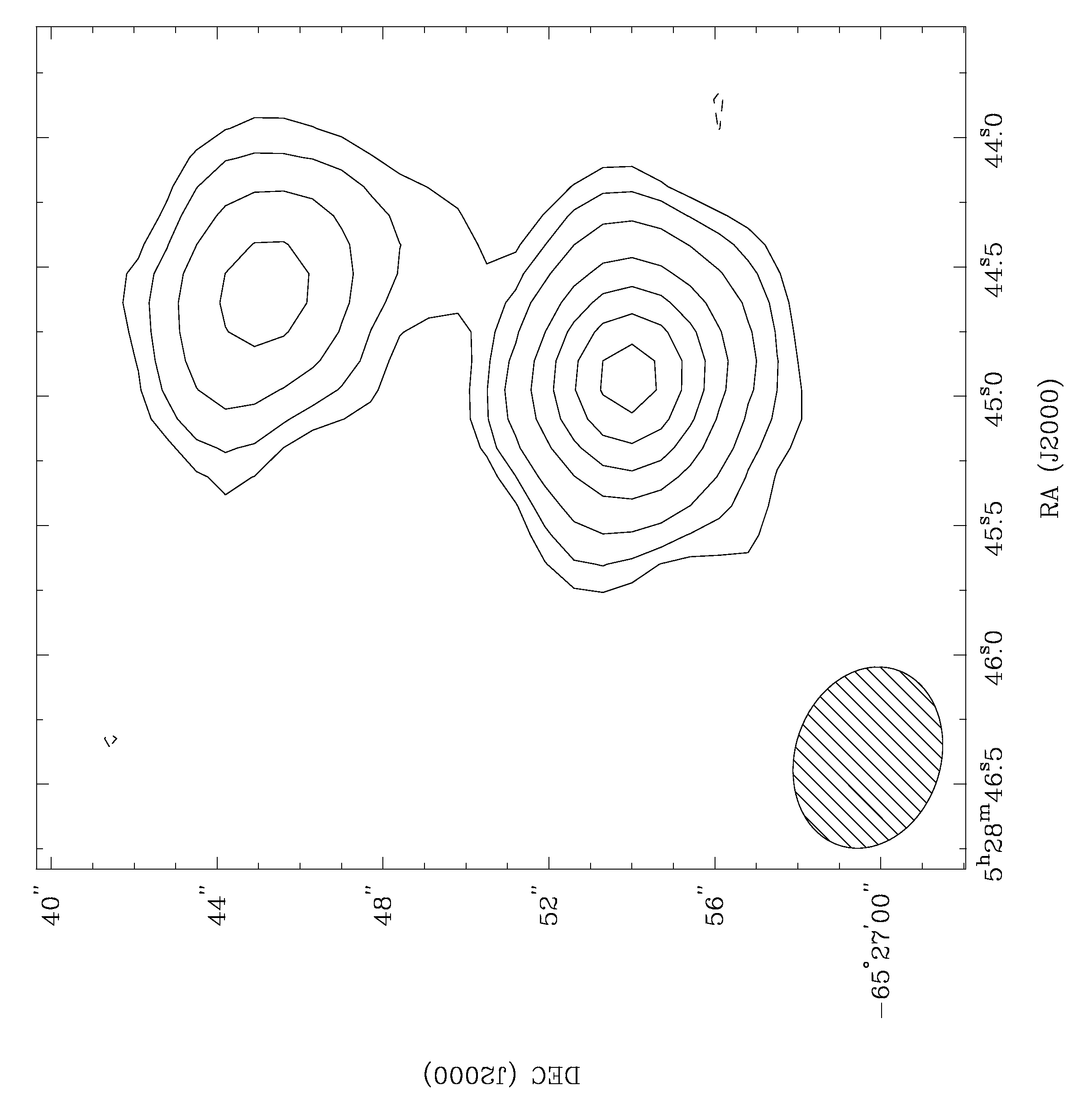} &
\includegraphics[scale=0.40,angle=-90]{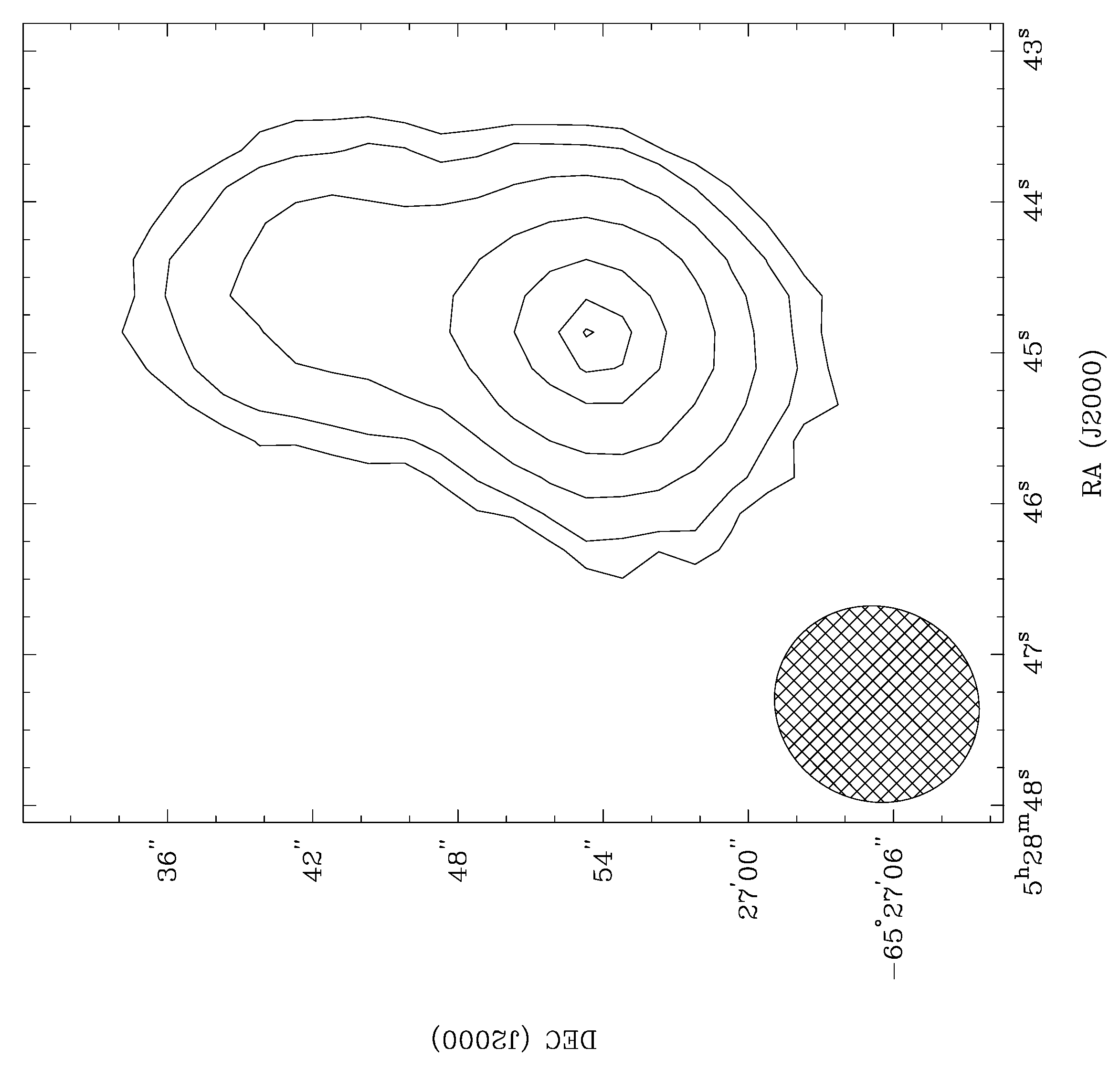} 
\end{array}$
\caption{
Robustly weighted contour maps of AB Dor A and B (northern component), using the uv data from 15 baselines. The maps on the left are for 21 Nov 2006; on the right 
8 Jan 2007.  The 8.64 GHz results are at the top; 4.80 GHz below.
Map (a) has highest and lowest contour levels of 2.62 and 0.20 mJy/beam, and the system noise 
over a large clear area around the source is 51 $\mu$Jy/beam. The beam axes are 3.36 and 2.84 arcsec at PA 71.2 deg.
Map (b) has highest and lowest contour levels of 3.37 and 0.15 mJy/beam, and corresponding noise 47 $\mu$Jy/beam. The beam axes are 4.45 and 3.51 arcsec; PA 72.0 deg.
Map (c) has corresponding contour levels of 2.85 and 0.25 mJy/beam and noise
65 $\mu$Jy/beam.  The beam axes are 4.82 and 4.52 arcsec; PA --23.6 deg.
Map (d) has corresponding contour levels of 4.08 and 0.33 mJy/beam  and noise 
 62 $\mu$Jy/beam. The beam axes are 8.56 and 8.01 arcsec; PA - 23.6 deg.}
\label{AB_Dor_contour_map} 
\end{center}
\end{figure*}

Ideally, we require an array configuration providing angular resolution much less than the $\sim$8~arc~sec separation of AB Dor A from its M3.5V visual companion at the four frequencies of interest, in order to identify the individual contributions of the two stars.
 In the event, we were granted a 1.5~km configuration (5 antennae) on the Nov 21/22 allocation, and a 750~m configuration on the Jan 8/9 allocation. Both these configurations (especially the 750~m one) were less than optimal for the task, but we were able to obtain satisfactory resolution of the stars' emissions at 4.800 and 8.640~GHz.   Reductions have been performed using the MIRIAD software package (Sault et al., 1995), which has become well-adapted for application to ATCA data. 

  The sampling interval for our observations was 10s and samples were accumulated  for 14.2 h in Nov, 2006 and 11.4 h in Jan, 2007. The two pairs of bands were alternately recorded in 25 m integrations and calibrated using MIRIAD. After we mapped the two sets and discovered evidence for correlated activity, we further divided each 25 m integration in the C/X bands into 5 m integrations in a search for finer temporal structure, and to assist in deriving  accurate correlation functions.
Analysis of the 1.384 and 2.368~GHz data was further complicated by the presence of persistent, strong interference, particularly on the ten shorter baselines. We found it, eventually, impracticable to produce useful light curves at the longer wavelengths, although resolved images were obtained at the longer wavelengths from integrations over the total allocation intervals.

  Lack of angular resolution with the more compact array of Jan 8, 2007 meant that the usual mapping tasks in MIRIAD gave problems for the separation of AB Dor A and B on all short time integrations.  Maps weighted with the `robust' option show an almost complete resolution at X band, however, and a usable result at C band.
 It was decided to use this option to produce all four maps shown in Fig~5. The resolution is then good enough to permit UVFIT to converge to consistent flux and rms values. The outer contour for the Jan 8 data (especially at X-band) shows the wavy structure compatible with a significant gap in uv coverage between the 5 compact baselines and antenna 6.  
   To attain better resolution at L/S on both observing days we tried the 5 longest baselines to antenna 6, resulting in a marked reduction in the S/N of the 25 minute integrations. Our L~band data were then found still contaminated by terrestrial interference and we did not analyse them further.  We also do not present the S~band light curve in this paper, since it is not in the same S/N category as the shorter wavelength data.

 We used the MIRIAD task UVFIT on both 25-m and 5-m integrations for the C and X bands. UVFIT fits a model to the uv data, using accurate offsets from the phase centre for the A and B components.  These offsets were derived from cleaned contour maps of the 5-baseline data to antenna 6, using the task IMFIT. Our final weighted means of the four independent sets of offsets at each frequency are $\Delta x = 0.430$ arcsec and -–1.797 arcsec; and $\Delta y = +9.865$ and +18.625 arcsec for components A and B, respectively. The errors in these measurements are $\pm$0.035 and $\pm$0.065 arcsec, respectively.
These offsets were used in deriving the flux densities.

We have made naturally-weighted contour maps at 4.80 and 8.64~GHz from all the 25~min.\ integrations available on each of the three observation periods (two in November 2006 and one in January 2007). Maps were made at each of the following combinations of baselines; (1) the total of 15 baselines; (2) the 10 compact baselines; and (3) the 5 long baselines. A comparison of the maps allowed us to select  the combination that would permit measurement of the flux densities from the individual 25~min.\ integrations with the best combination of sensitivity and angular resolution.   The best images allowed us to check the used angular offsets of the stellar images from the field centre  (see Figure~5).  

The radio position corresponding to the best integrated image yields AB Dor A at 2006.894 to have  RA: 05 h 28 m 44.924 s +/- 0.007; and  dec: 
--65 deg 26 arcmin 53.910 arcsec +/- 0.039.  This may be compared with the ICRS position for 2000.0, i.e.
RA: 05 h 28 m 44.8280 s +/- 0.001; and  dec: --65 deg 26 arcmin 54.853 arcsec +/- 0.004.  Mean proper motions that could be derived from this pair of positions and 6.894 y interval are different from the HIPPARCOS-based (van Leeuwen, 2007) values listed by SIMBAD, however. This may well be related to orbital motion within this multiple system, as noted in a similar context by Martin and Brandner (1995), from whose data over the 65 year interval 1929-94 proper motions of +51$\pm 5$ and +130$\pm 5$ mas y$^{-1}$ in RA and dec respectively can be found.  These are closer to the +87$\pm 3$ and  and +137$\pm 3$ results of our measures, though still well outside the measurement errors from the SIMBAD figures of 33.2$\pm 0.4$ and 150.8$\pm 0.7$.  But, from the orbit of the AB Dor A-C system shown by Nielsen et al.\ (2005), it is clear that
orbit-related proper motion changes of the order of tens of mas y$^{-1}$ can be reasonably expected from year to year.
The short-interval HIPPARCOS values may thus be significantly affected in this way.


Maps of the source are shown in Figure~\ref{AB_Dor_contour_map}. A log of the observations and the resulting mean flux densities for each night are given in Table~\ref{table:statistics}, which also contains the mean spectral indices
$\alpha$, with the spectral index defined by S($\nu$) = C$\nu^{\alpha}$.  The polarized flux in this table 
refers to the circular (V) component.

										   
\begin{table*}
{\scriptsize
\centering
\begin{minipage}{100mm}
\caption{ATCA observation log and summary of data.}
\label{table:statistics}
\begin{tabular}{ @{}ccccccccc@{} }
\hline				   
Date & Dur.\ & Star & Freq.\ &Mean  &Pol'ized&                                                        \multicolumn{3}{c}{\mbox Spectral index  }\\		   
Config. & hr& & GHz	 & Total & Frac.\ & $\alpha^{1}$: & $\alpha^{2}$: & $\alpha^{1}-\alpha^{2}$\\
 & & & & Flux  & V/I & 2.37--4.80 & 4.80--8.64 & \\
& & & & (mJy)  & & GHz & GHz & \\
\hline
 & & & & & & & & \\
21 Nov. 2006& 17.1 & AB Dor A &2.37 &3.20 (0.11)& 0.04 (0.10) &+0.27 (0.05) &--0.35 (0.03)  & 0.62 (0.06)  \\	   
1.5B & & & 4.80 & 3.88 (0.04) & 0.05 (0.06) & & &	\\	   
& & & 8.64 & 3.15 (0.05)  & 0.15 (0.07)	& & & \\
&&&&&&&&\\
& & AB Dor B & 2.37 & 1.60 (0.11) & 0.33 (0.10) &+0.17 (0.10) & --0.22 (0.06) & 0.39 (0.12) \\
& & & 4.80 & 1.80 (0.04)  & 0.30 (0.06) & & & \\

& & & 8.64& 1.58 (0.05)& 0.34 (0.07) & & & \\
\hline
 & & & & & & & & \\
08 Jan. 2007 & 13.3 & AB Dor A & 2.37 & 3.65 (0.11) & 0.01 (0.16) &+0.12 (0.05) &--0.20 (0.03) &0.32 (0.06)  \\
750A& & & 4.80 & 3.96 (0.05)  & 0.07 (0.07) & & &  \\
& & & 8.64 & 3.51 (0.05) & 0.12 (0.07) & & & \\
&&&&&&&&\\
& & AB Dor B & 2.37& 1.73 (0.11)  & 0.17 (0.16) &--0.21 (0.10) &+0.18 (0.08)  & --0.39 (0.13) \\
& & & 4.80 & 1.49 (0.05)  & 0.25 (0.07)& & & \\
& & & 8.64 & 1.66 (0.05)  & 0.25 (0.07) & & & \\
\hline
\end{tabular}
\end{minipage}
}
\end{table*}

\subsubsection{Flux Density Plots}

\begin{figure*}
\vspace{0cm}
\begin{center}$
\begin{array}{cc}
\hspace{-0.25cm}
\includegraphics[scale = 0.29, angle = -90]{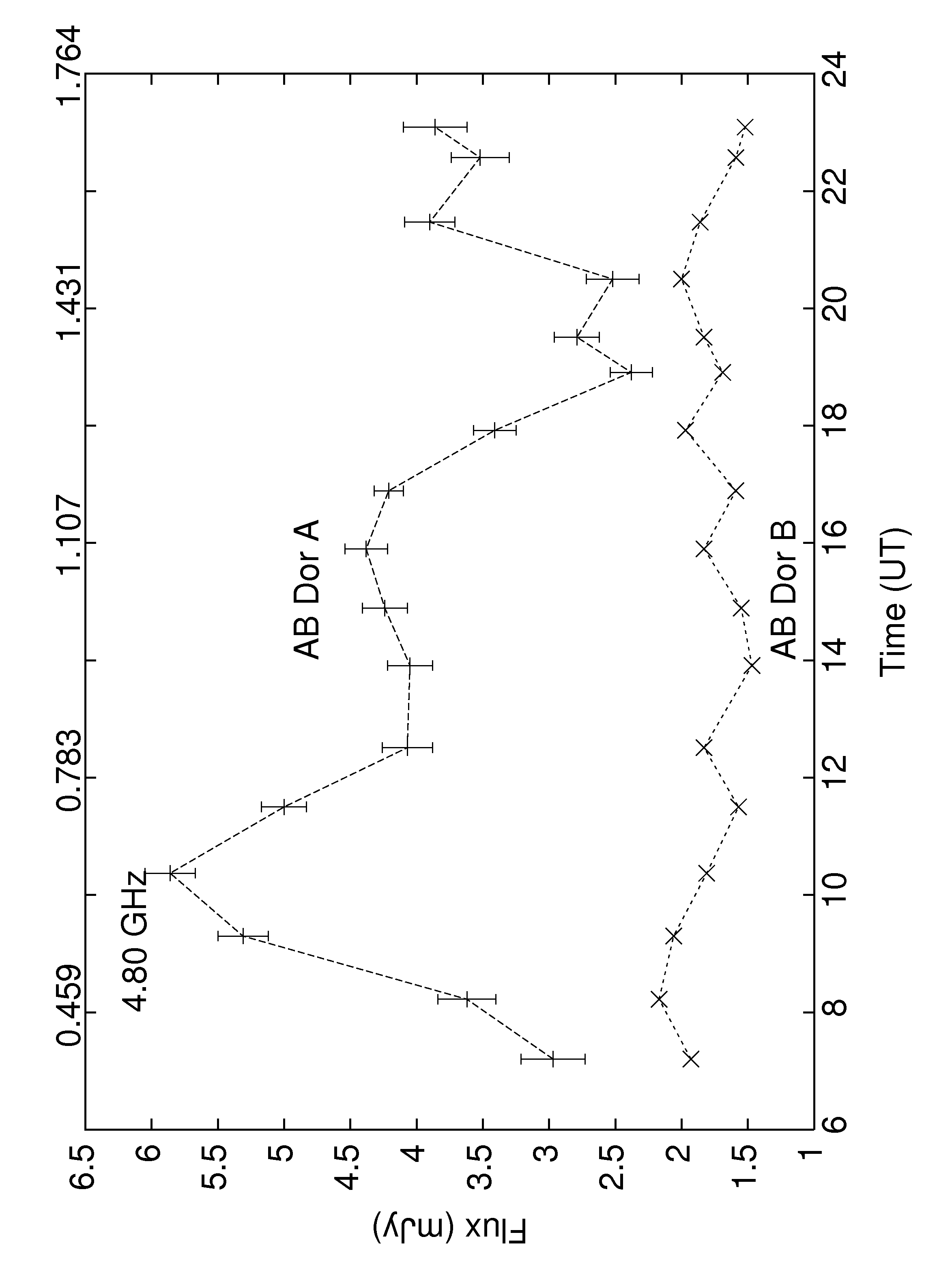} &
\includegraphics[scale = 0.29, angle = -90]{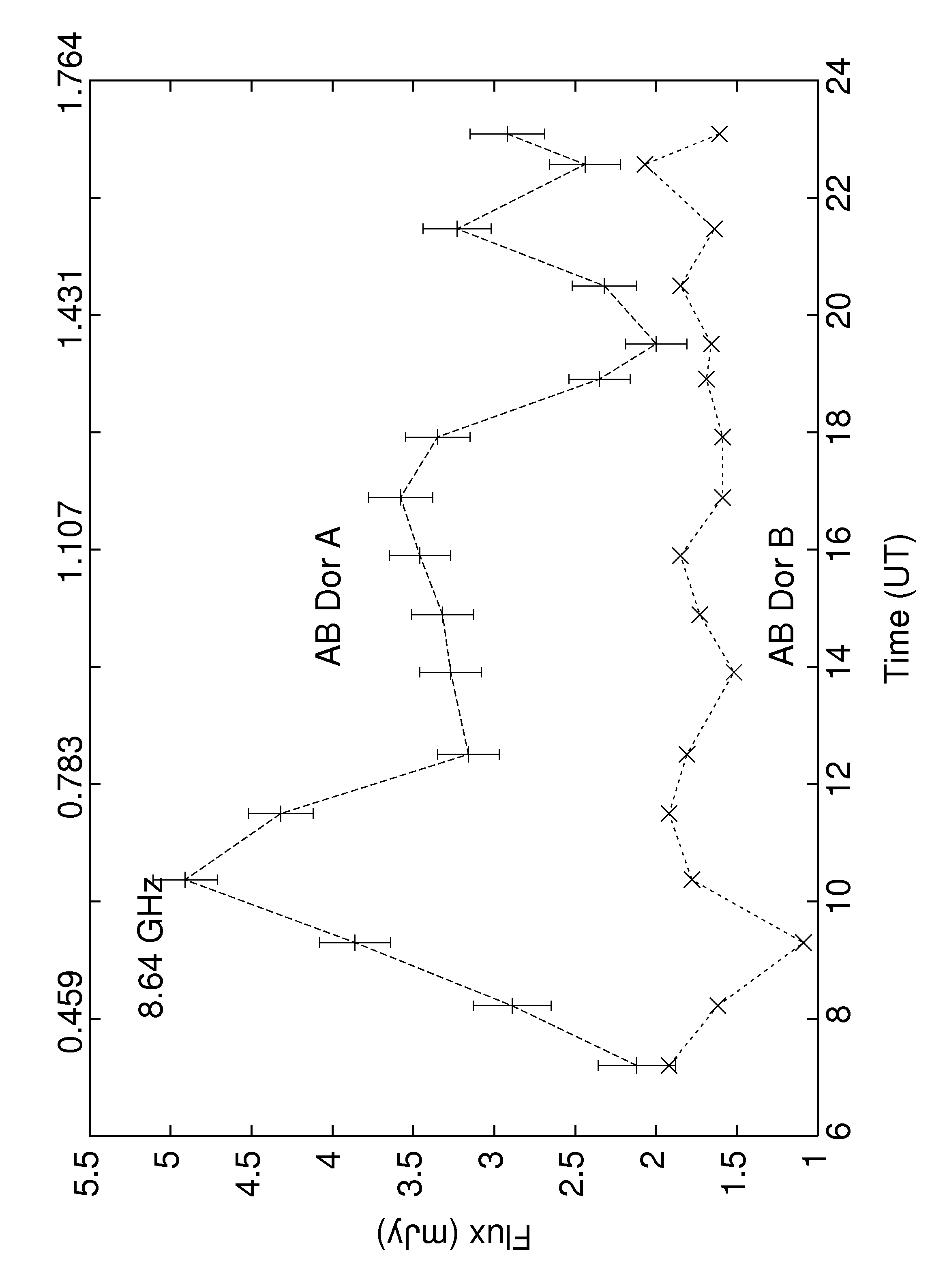} \\ 
\hspace{-0.25cm}
\includegraphics[scale = 0.29, angle = -90]{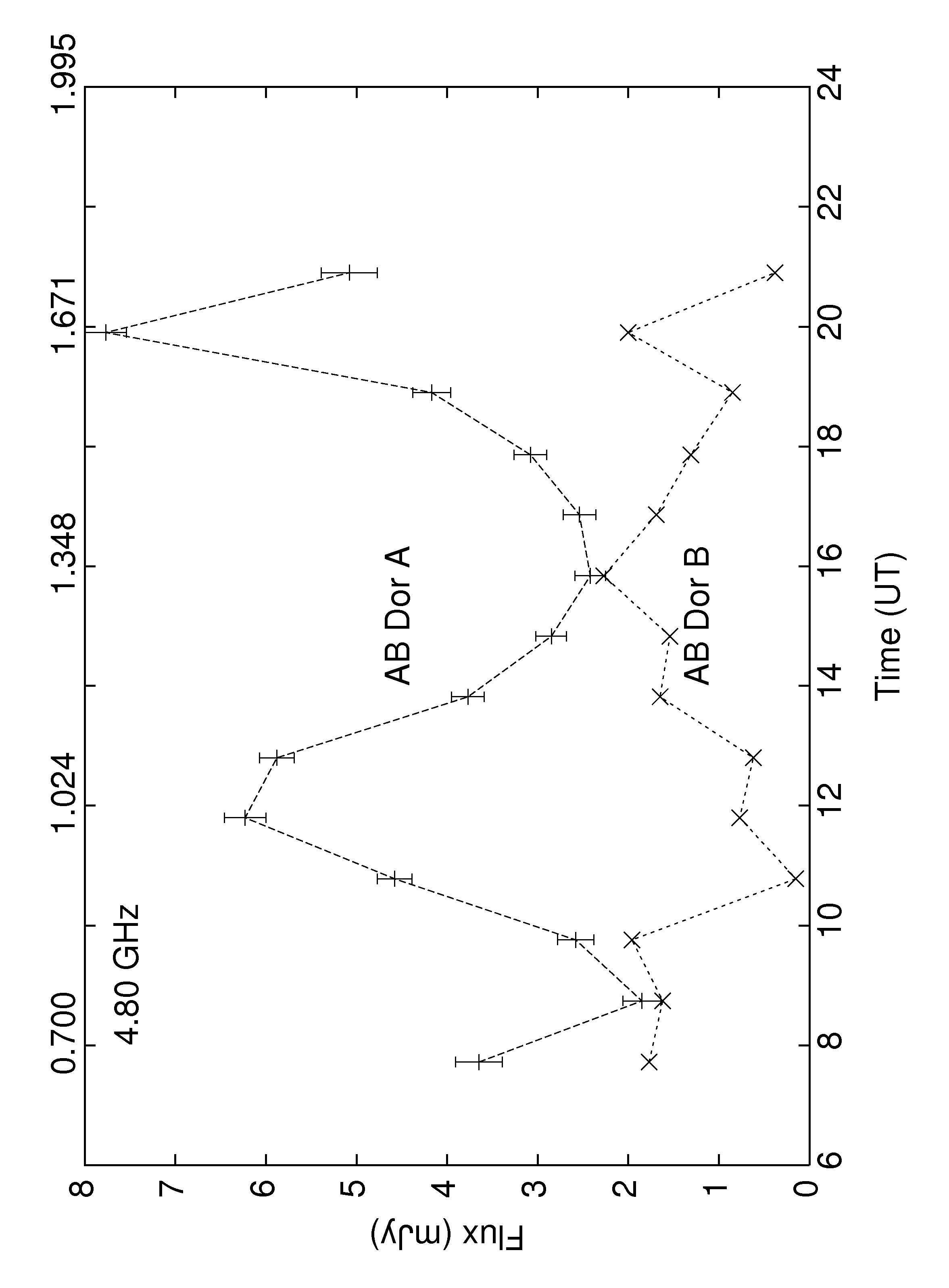} &
\includegraphics[scale = 0.29, angle = -90]{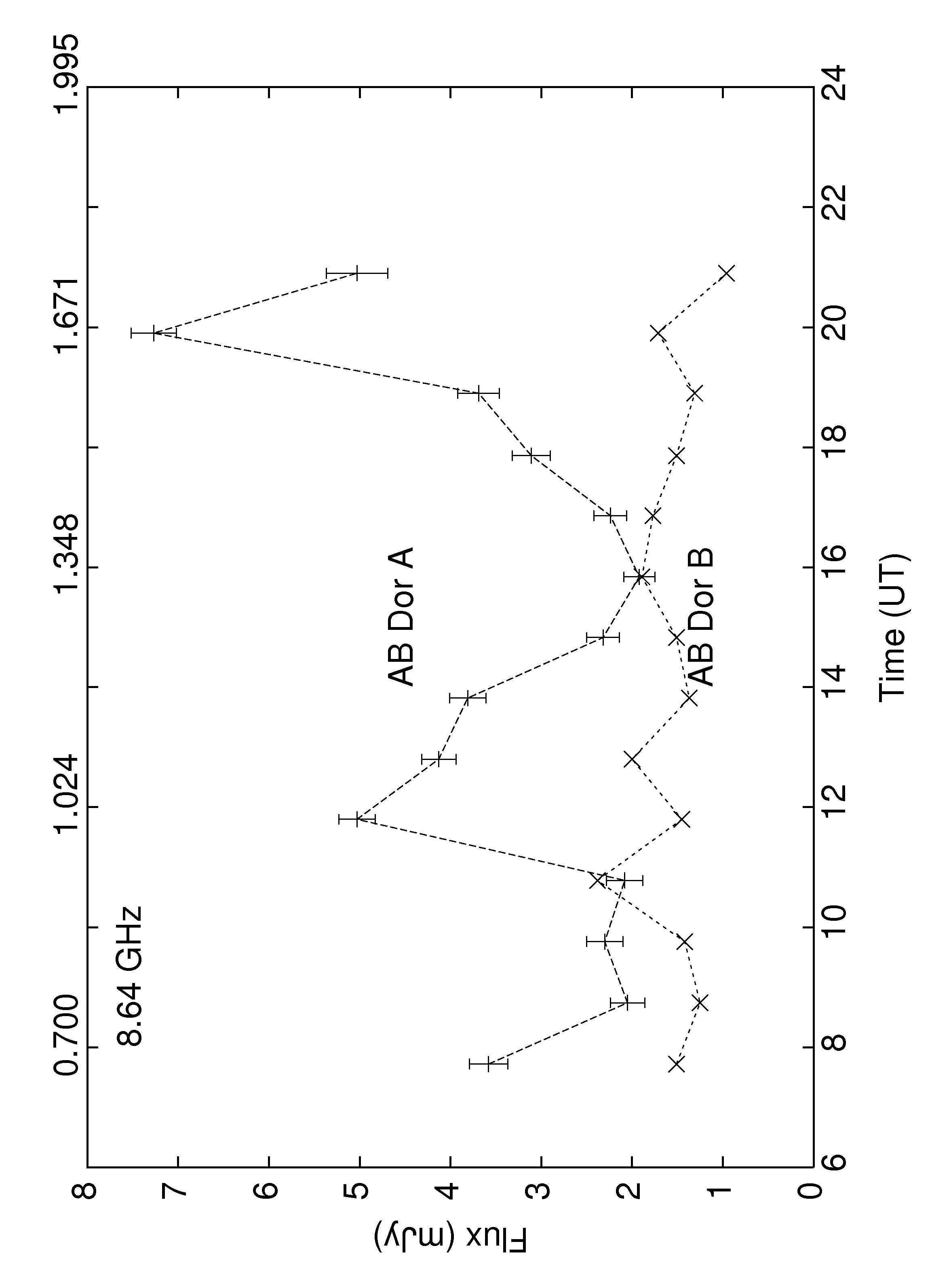} 
\end{array}$
\end{center}
\vspace{-0.4cm}
\caption{Plots of flux densities corresponding to the midpoints of 25~min.\ integrations on 2006 Nov 21 (upper panels)  and 2007 Jan  08 (lower panels) for AB Dor A and B. These points utilize data from all 15 baselines
at 4.80 and 8.64 GHz.  Photometric phases, corresponding to the times shown on the lower axis, are displayed on the top axis, using Innis et al.'s (1988) ephemeris. The rms error measures for both components as derived by UVFIT were the same; they are omitted from the lower light curves to avoid possible confusion when the signals are close.}
\label{AB_Dor_flux_densities}
\end{figure*}

The flux densities derived with UVFIT for the AB Dor system are plotted in the montage of Figure~6  for the observing runs on 2006 Nov 21 and 2007 January 08. The date, frequencies and stars are identified in the panels. Each point in these plots represents the flux density at the mid-UT of a 25 min integration, and the error bars represents the rms residual of the fit. Clear, long-lasting increases for AB Dor A are seen on both dates, soon after the start of observations on Nov 21 and a few hours in and again towards the end of the run on Jan 8.
The 2.27 GHz data-set is suggestive of correspondence with these effects,
but it is much less clear against the relatively high scale of noise
for the reasons given before.


Table~\ref{table:flares} presents estimates of peak flare intensities (i.e. excess emission above the background) and their total durations at both higher frequencies. The angular resolution was sufficient to show that the M-type dwarf binary, AB Dor B, remained relatively quiescent on all three nights and it is clear that the fluxes measured for AB Dor A are not significantly influenced by the companion. The scale of noise in the measurements is provided by both UVFLUX and UVFIT.  The relatively strong cross-correlation of the C and X-band data, discussed later, provides further confirmation of the reality of the signal excursions during flaring episodes. Base levels for these events can be reasonably set at 2.0 mJy for all the curves, except for the 4.80 GHz curve on Nov 21, when the base level appears closer to 2.5  mJy. The excess flare fluxes are based on these levels.

\begin{table*}
\vspace{-0.3cm}
{\scriptsize
\centering
\begin{minipage}{100mm}
\caption{Microwave flare statistics.}
\label{table:flares}
\begin{tabular}{ @{}ccccccccc@{} }
\hline
Date & Flare & Flare & \multicolumn{3}{c}{\mbox Peak Flux Density (mJy)}& \multicolumn{3}{c}{\mbox Spectral Index}\\		   
& & UT (h) & 2.37~GHz	 & 4.80~GHz & 8.64~GHz  & $\alpha^{1}$: & $\alpha^{2}$: & $\alpha^{1}-\alpha^{2}$\\
 & No & Duration & &  & & 2.37--4.80 & 4.80--8.64 & \\
& & &   &  & & GHz & GHz & \\
\hline
 &  &  &  &  &  &  &  & \\
06 Nov. 21 & 1& 07:25--14:50&2.33 (0.49)& 3.24 (0.20)& 2.71 (0.21)& 0.47 (0.31)& --0.30 (0.17)& 0.77 (0.35)\\
& 2& 12:50--18:50& 1.06 (0.38)& 1.68 (0.16)& 1.37 (0.16)& 0.65 (0.53)& --0.35 (0.26)& 1.00 (0.59)\\
& 3& 20:00--$\ge$23:00& $\le$0.5 & 1.11 (0.16) & 0.69 (0.16) & $\ge$1.46 & --0.81 (0.47)& $\ge$ 0.67 (0.70)\\  
&&&&&&&&\\
07 Jan. 08 & 1&10:00--15:50 &1.72 (0.34) & 4.23 (0.21) & 3.07 (0.22) & 1.28 (0.29) & --0.55 (0.15) & 1.83 (0.33) \\
& 2 & 16:50--$\ge$21:50& 2.48 (0.40) & 5.95 (0.26) & 5.31 (0.26) & 1.24 (0.24) & --0.19 (0.11) & 1.43 (0.26) \\
\hline
\end{tabular}
\end{minipage}
}
\end{table*}

\subsection{Optical Observations}

The broadband $B$ and $V$ photometry and high resolution, wide-range \'{e}chelle spectroscopy components of this campaign were discussed in more detail in Innis et al.\ (2008) and Budding et al.\ (2009).  The latter paper presented an empirical model of surface activity, generally contemporaneous with the radio and X-ray data, consisting of four main centres.   That paper
also gave reasons why surface features of the scale in question could be considered stable over timescales of a few months, in general.  
One of the optical features discerned (a) appears to consist of a strong umbral region near, but somewhat above, phase zero ($\phi = 0.03$), that may also be related to a chromospheric enhancement at phase 0.95.  This location agreement
strengthens the argument of applicability of not exactly time-coincident data to models across the complete spectrum (see also Lim et al., 1994). 
The other three activity centres (b, c and d) correspond essentially to chromospheric enhancements, defined particularly in high dispersion Ca II emission K lines, are centred around phases (0.26, 0.58 and 0.84) respectively.  We shall refer to this representation in interpreting the X-ray and radio coronal emission.

A more detailed combination of almost simultaneous optical and X-ray observations of AB Dor using the SemelPol polarimeter and UCLES spectrograph at the AAT and the Chandra satellite over a five day interval in late 2002
was carried out by Hussain et al (2007). It is of interest that there were qualitatively similar results regarding the  small number of longitudinally confined major centres of activity at high latitudes.  These connect optically identified photospheric and chromospheric features, projected magnetic field strengths and X-ray emission levels, with the different concentrations having separations of order 90$^{\circ}$.  This configuration is 
frequently found together with a large polar spot, and
appears characteristic of this type of active cool dwarf star (Hussain et al., 2007).

\section{Detailed Data Analysis}
\subsection{Radio Spectral indices}
\label{indices}
The spectral indices between 2.37 and 4.80~GHz and between 4.80 and 8.64~GHz are listed in separate columns in Tables 1 and 2. These are computed assuming a power-law spectrum as defined in Section 2.2.1.

The indices in Table~\ref{table:statistics} are those obtained from the averaged flux densities for each observing session's data, including both the background and flare emission. The error assigned to each spectral index and their differences is that obtained by propagating the bracketed errors assigned to the appropriate flux densities and spectral indices (cf.\ Bevington, 1969). The final column provides a test of the significance of the differences between the pairs of spectral indices. For AB~Dor the spectral indices appear to change from positive to negative near a frequency of 4.8~GHz, indicating that the frequency at which the radio corona changes from being generally optically thick to thin is not far from this frequency.  This effect is also apparent in AB Dor B on two of the three nights, and while unconfirmed for the last run the low S/N of the
data should be kept in mind.

Table~\ref{table:flares} identifies the flares in AB Dor's emission and lists their total durations; in addition, the excess peak flux density above an estimated background emission is given at each frequency. The error assigned to the excess peak flux density is greater than that in Table~\ref{table:statistics} because the difference involves two measures, each of which has its own error. Nevertheless, the spectral index is positive below, and negative above, $\sim$4.80~GHz. The significance of this change is determined from the last column; the relevant quantity being found with an overall significance that ranges between 1.7 and 5.5 sigma.   The spectral turnover result can be considered as essentially well-established from the higher frequency ranges, however. The relatively lower significance of results involving the S-band integrations has been discussed above, and does not change the main finding about the spectral index.
Weighted means of $\alpha^{1}$ and  $\alpha^{2}$ (weights $\sigma^{-2}$) yield $<\alpha^{1}>$~=~$+$1.02$\pm$0.15 and $<\alpha^{2}>$~=~$-$~0.33$\pm$0.07.   

\subsection{Polarization}
\label{polarisation}

  We tested the circular polarization fractions for AB Dor A and B listed in Table 1 for 21 Nov 2006 by making contour plots at 4.80 and 8.64 GHz 
in Stokes V, using all uv data for that date.
A comparison between the V/I results in Table 1 and the two contour maps shown in Figure~7 shows they are in reasonable agreement. 
At 4.80 GHz, Table 1 shows that AB Dor A has a significance level of 1-$\sigma$, consistent with its weak V contours. 
Component AB Dor B has much more intense V contours, consistent with the 5-$\sigma$ significance in Table 1.
  We then examined the V/I values for the 4.80 GHz data of 08 Jan 2007. The 13 twenty-five min integrations that form its radio light curve 
in Figure~6 were sub-divided into 65 five-min integrations in order to search for episodes of ECME, 
which can have  short durations in late-type stars (Slee et al. 2008). 
This analysis yielded two interesting conclusions:
(1) 4 of the 5-min integrations (spread over $\sim$10 h) showed significant polarization fractions for component B with $\sigma$-values 
of 3, 4, 7, 9.
(2) None of the polarization fractions for component A had $\sigma$-values exceeding 0.8, and most  were much lower.

   
\begin{figure*}
\begin{center} $
\begin{array} {cc}
\hspace{-0.8cm}
\includegraphics[scale=0.35, angle=-90]{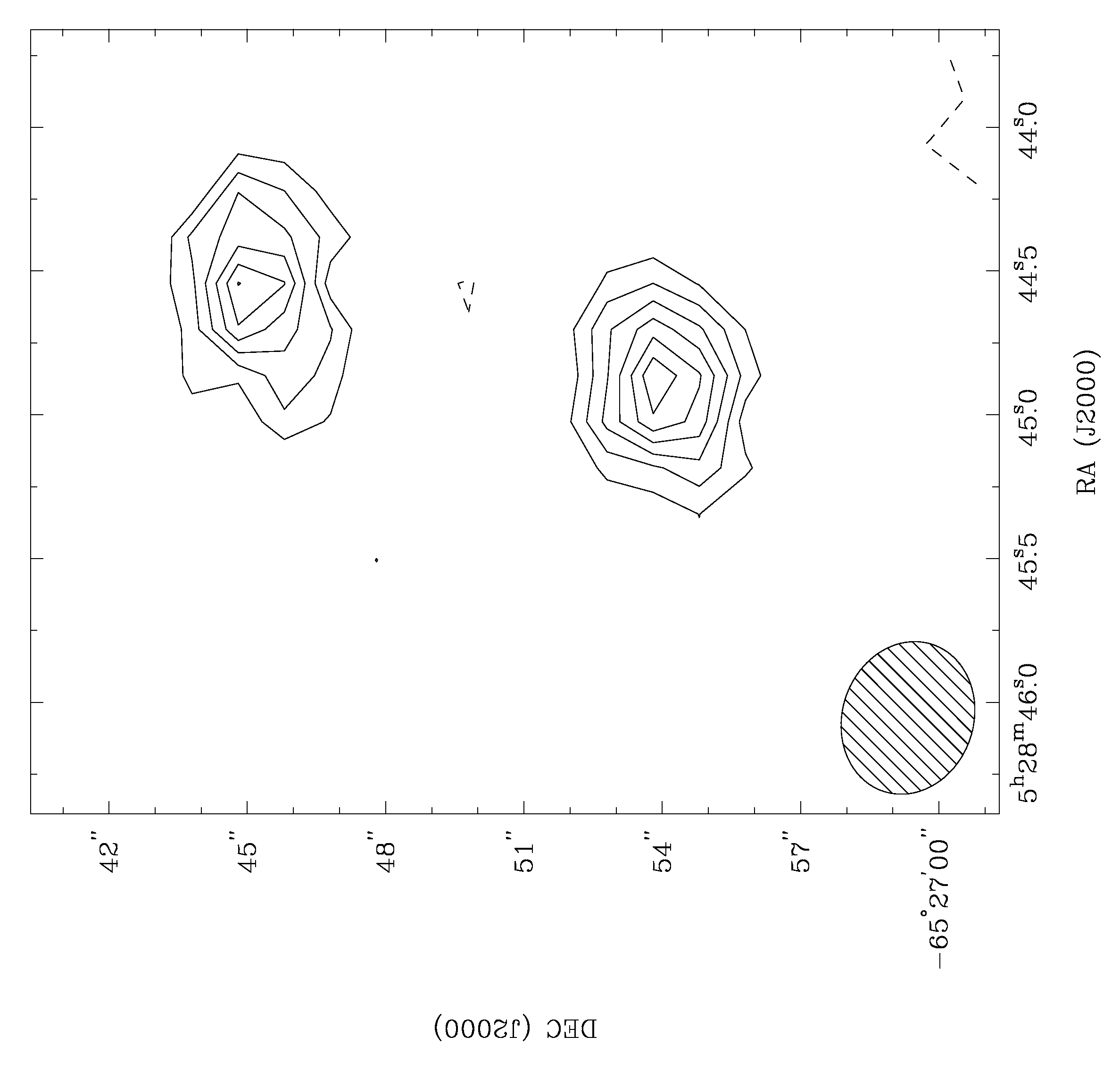} &
\includegraphics[scale=0.35, angle=-90]{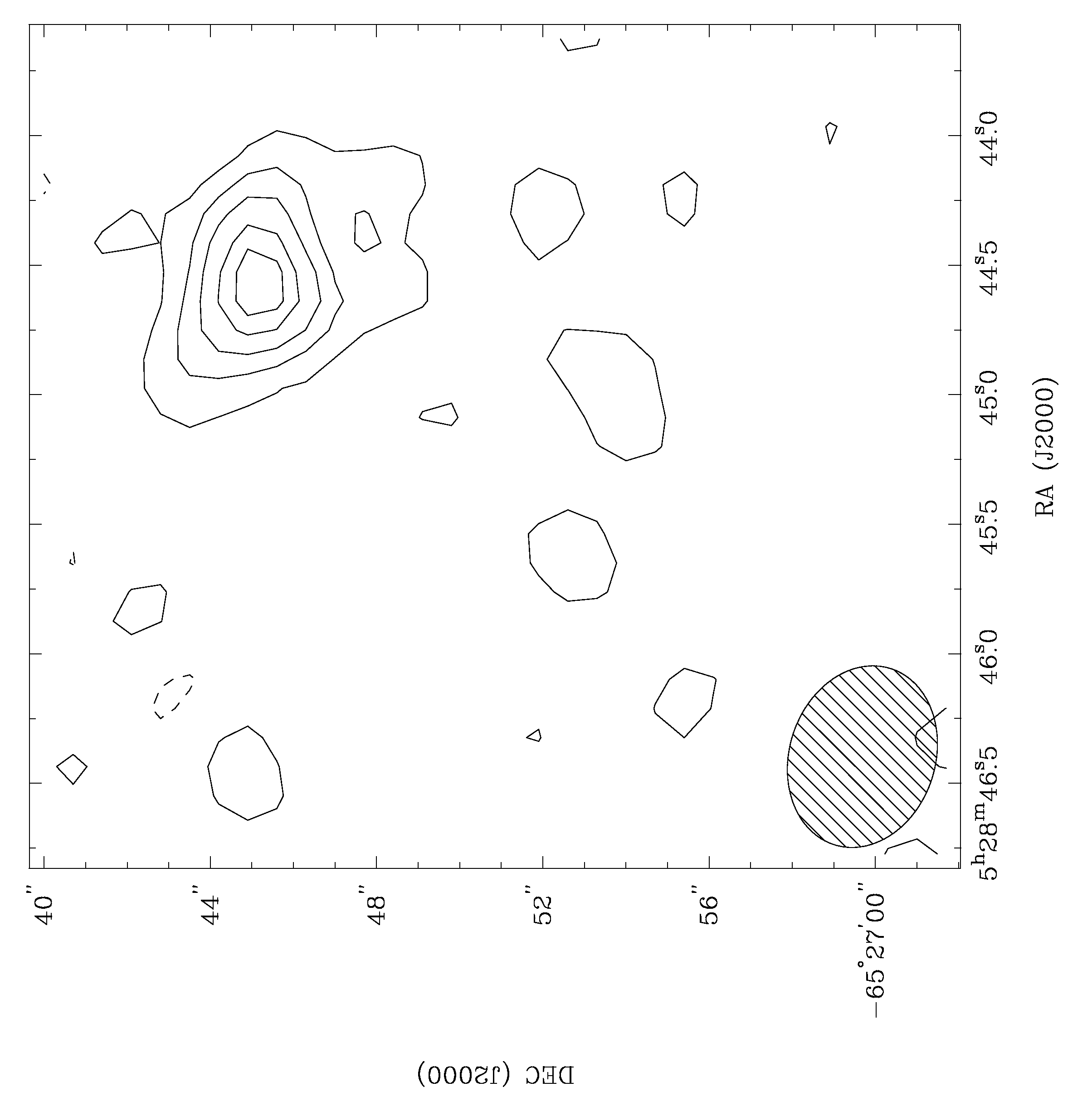} 
\end{array}$
\caption{Contour maps of AB Dor A and B (northern component) in the Stokes V polarization at 8.64~GHz (left) and 4.80~GHz (right), using all the data from 15 baselines on 2006 November 21 and robust weighting. The highest and lowest contour levels at  8.64~GHz are 2.299 and 0.120~mJy~beam$^{-1}$ respectively. The rms system noise level is 47~$\mu$Jy~beam$^{-1}$ and the FWHM (full width at half maximum) of the restoring beam (lower left) is 3.36$\times$2.84~arcsec, with major axis in PA = 71.2$^{\circ}$. At 4.80~GHz, the maximum and minimum contour levels are 0.423 and 0.086~mJy~beam$^{-1}$ respectively,
 and the rms system noise level is 39~$\mu$Jy~beam$^{-1}$. The FWHM of the restoring beam 
 is 4.45$\times$3.51 arcsec, with major axis in PA 72.0$^{\circ}$.}
\label{figure_7} 
\end{center}
\end{figure*}
 

 There is an indication that the frequency dependence in spectral index of AB Dor B changes sign from 
2006 November 21 to 2007 January 08. In that regard it differs from the more
consistent spectral behaviour of AB Dor A.
It would be useful to try to check the behaviour of AB Dor B
with VLBI observations.

\subsection{Flares}
\label{flare_power}
In order to estimate the radio power emitted during the flares listed in Table~\ref{table:flares} we need a spectral model.
For this purpose, we used the 5-baseline-data at 2.37, 4.80, and also 8.64 GHz, if a thoroughly calibrated contour plot showed an image well above the system noise and defects produced by incomplete sampling of the $uv$-plane. These high angular resolution data provided accurate off-sets from the phase centre for AB Dor A and B.  At 2.37 GHz, the 5 baseline data provided reliable (though noisy) mean values for the flux densities over the two long runs. This gave rise to our two estimates of the  spectral indices between 2.37 and 4.80 GHz listed in Tables 1 and 2.

From this, we propose the following spectral model over the frequency range from 10 MHz to 300~GHz:

\begin{itemize}
\item (a)	S($\nu$) = A$\nu^{1.02}$  between 0.01 and 4.80~GHz
\item (b)	S($\nu$) = B$\nu^{-0.33}$ between 4.80 and 20~GHz
\item (c)	S($\nu$) = C$\nu^{-1.5}$  between 20 and 300~GHz
\end{itemize}
where S($\nu$) is the flux density and A and B are constants that can be evaluated by substituting in the appropriate equation a measured value of flux density at a particular frequency.  For the highest frequency range we have no data directly available, but we can tailor the observed result for the middle range of frequencies into a feasible high-frequency fall-off, using the known properties of solar flares as a guide (e.g.\ Hurford, 1986). 
Using the 4.80~GHz peak flux density in Table~\ref{table:flares} of 5.95$\pm$0.26~mJy in flare No. 2 on 2007 January 08 as the reference, we find A = 1.20, B = 9.99, C = 332.  Examination of the errors of the flux measures at the respective frequencies show A can be estimated to within $\sim35$\% of its real value, while B should be accurate to within
$\sim10$\% for a given flare. The high frequency extension illustrates the point that a considerable amount of flare radiation is likely beyond what is typically considered
in microwave flux calculations.

Integrating over the three frequency ranges with this model, we find the peak power emitted by flare 2 on January 08 is 5.12$\times$10$^{18}$~W. Here we use a distance to AB Dor of 14.9~pc.  Table~\ref{table:flares} indicates that a typical flare duration to half--brightness is about 3~hours, yielding a total  energy output of 5.5$\times$10$^{22}$~J. Since the peak flux densities of the identified flares in Table~\ref{table:flares} vary by about a factor of 5-7 at 4.80 and 8.64~GHz, but their durations appear similar, their total energy outputs are likely to vary by similar proportions.

These estimates of peak and total powers are necessarily uncertain, because we have direct values of the spectral indices only over a small frequency range and, even then, the averaged indices $\alpha^{1}$ and $\alpha^{2}$ have uncertainties of around 15 and 20\% respectively.  But the uncertainties at the low and intermediate frequencies have relatively small weight on the total result: it is in the range above 20~GHz that there may be significant departures from the radio flare energies typically considered. Even with the comparatively steep spectral index of --1.5, this region would contribute more than half of the total radio emission.

\subsection{Intensity structure}
\label{ISC}

\begin{figure}
\hspace{-0.5cm}
\includegraphics[scale=1, angle=0]{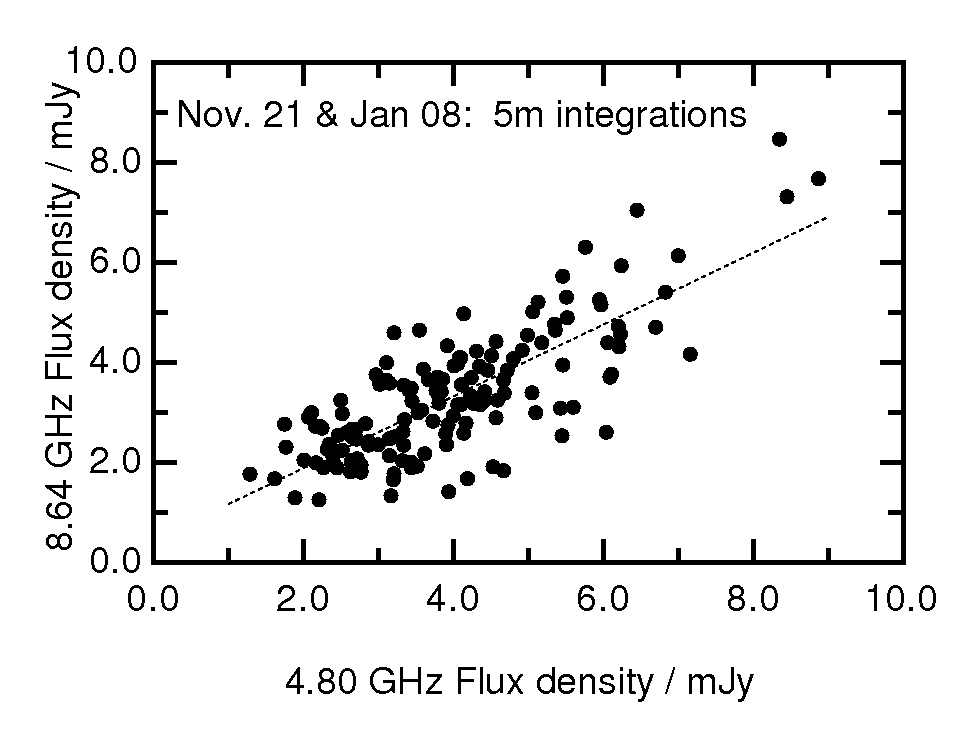}
\caption{Correlation between 153 five--minute integrations of the AB Dor data at 4.80 and 8.64~GHz on 2006 November 21 and 2007 January 08. The regression line has a slope of 0.72$\pm$0.05 and the correlation co-efficient of 0.78 can
be regarded as highly significant for a 153 point sample. The individual points in the plot are formed from flux densities with rms residuals of $\sim$0.4~mJy.}
\label{AB_Dor_correlations}
\end{figure}

A good idea of the temporal intensity structure in the radio measurements can be gained by plotting a scatter diagram of the data after  subdivision into 5--min.\ integrations. Figure~\ref{AB_Dor_correlations} shows the 8.64 GHz 5-min. integrations plotted against those at 4.80 GHz for the long observing periods on November 21 and January 08. A linear regression line with a slope of 0.72$\pm$0.05 has been fitted to the 153 points in the plot; the correlation coefficient is 0.78.

Figure~\ref{AB_Dor_correlations} confirms that most of the intensity variability is in the range 2--7~mJy. The means of
the five minute integrations and their standard deviations at 4.80 and 8.64~GHz are 3.96$\pm$1.43 and 3.30$\pm$1.32 respectively.  Typical intensity variability about these levels is thus in the range 2.6--2.8~mJy at the two frequencies.  Individual intensities for each point in Figure~8 have measurement errors between 0.35 and 0.45~mJy.

Previous observations of active stars (including AB Dor) have been published by Budding et al.\ (2002), showing that the highest correlation between 4.80 and 8.64~GHz intensity variations occurred in a variety of active systems if the set of observations at one frequency is shifted in time with respect to the other set. In most cases, the cross-correlation functions were symmetrical about a shift consistent with flux variations at 4.80~GHz being later than those at 8.64~GHz.  The relatively high cross-correlation peaks found in such studies give a basic confirmation of the underlying reality
of the observed phenomena, as well as, in a more detailed way, allowing  physical inferences about the locations and
behaviour of the emerging wavefronts. Similar effects were reported by Osten et al.\ (2002) and discussed
in the solar context by Bastien et al.\ (1998).  The timescale of observed time-shifts between correlated events at different frequencies is consistent, in general, with the emission being propagated by magnetohydrodynamic waves through a highly energized, turbulent corona.

In the present experiment, we have used 5--min.\ integrations from the long observational series on 2006 November 21 and 2007 January 08 as shown in Figure~\ref{AB_Dor_correlations}. In order to compute the cross-correlation function, we shift the 8.64~GHz set of 5-min.\ integrations with respect to the fixed 4.80~GHz set, starting at zero shift and going to a maximum of +7 or +8 shifts.  At each shift, we fit a linear regression equation and note the correlation coefficient and its significance. The process is then reversed, with the 4.80~GHz set of 5-min.\ integrations being shifted with respect to the fixed 8.64~GHz set to form a negative set of shifts. The auto-correlation functions can be computed in a similar way, although in this case the process need not be reversed. The adopted auto-correlation function is the geometric mean of the 4.80 and 8.64~GHz functions. The cross-correlations are plotted and fitted with Gaussians, from which the shifts listed
in the caption to Fig~9 were determined.  These time shifts are significant at the 3-4 $\sigma$ level.
A good description of the applied technique that includes appropriate
discussion of significance tests was given by Moore (1958).
The use of the Gaussian fitting function, apart from its empirical validity in the region of the maximum, can be understood intuitively in terms
of the scattering effect of the turbulent medium on emerging radiation fronts, given their inferred spatial separation.

\begin{figure*}
\includegraphics[width=16.5cm, angle=0]{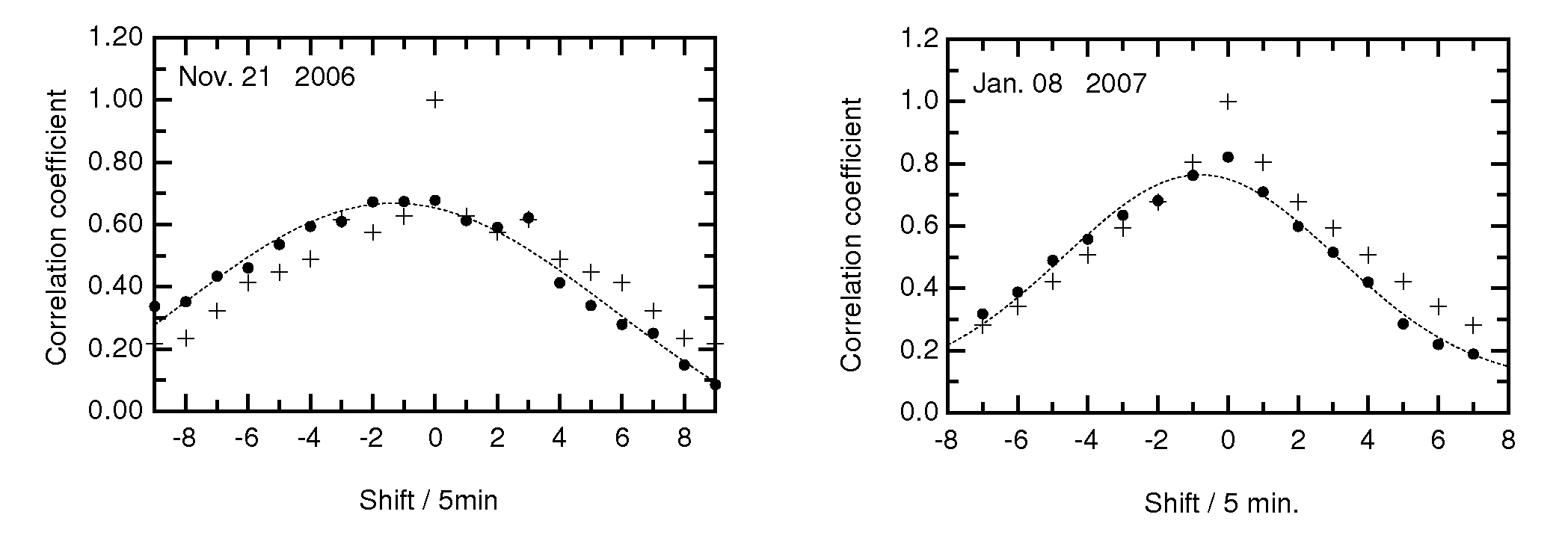}
\vspace{-0.6cm}
\caption{Correlation between 4.80 and~8.64 GHz intensity variations in two extensive sets of observations; each data-set has been subdivided into 5-minute integrations. The filled circles denote the cross-correlation function, which is smoothed by fitting a Gaussian. The crosses define the averaged 4.80 and 8.64~GHz auto-correlation function. The derived negative shifts of the cross with respect to the auto-correlation functions imply that 4.80~GHz intensity variations follow those at 8.64~GHz by 4.5$\pm$1.3~min on 2006 November 21 and 4.0$\pm$1.1~min on 2007 January 08. The procedure is described in Section~\ref{ISC} }.
\label{AB_Dor_correlations_48_86}
\end{figure*}

The resulting correlation functions are presented in Figure~\ref{AB_Dor_correlations_48_86}, which shows that the cross-correlations are systematically displaced in a negative direction with respect to the mean auto-correlations.
The trend of negative shifts for the Nov 21, 2006 data yielded a mean lag of --4.50$\pm$1.30 min.
The more regular cross-correlation function for the Jan 08, 2007 observations gave a mean shift of --4.00$\pm$1.13 min.
The lags should thus be regarded as confirmed at the 3$\sigma$ level.

 Similar findings were reported by Budding et al.\ (2002), whose correlation analysis of four long observations of AB Dor in 1994 and 1997 yielded shifts of 4.0 to 6.5~min. on three of the four nights (Table 1 in Budding et al., 2002).  Typical errors
for these determinations were given as $\sim2$ min, so while differences in the X to C-band shifts for individual
flares cannot be reliably specified, the shifts themselves can be regarded as well-confirmed with closely comparable results
obtained on 5 out of 6 independent data-sets.

\begin{figure}
\vspace{-0.3cm}
\includegraphics[width=7.2cm, angle=-90]{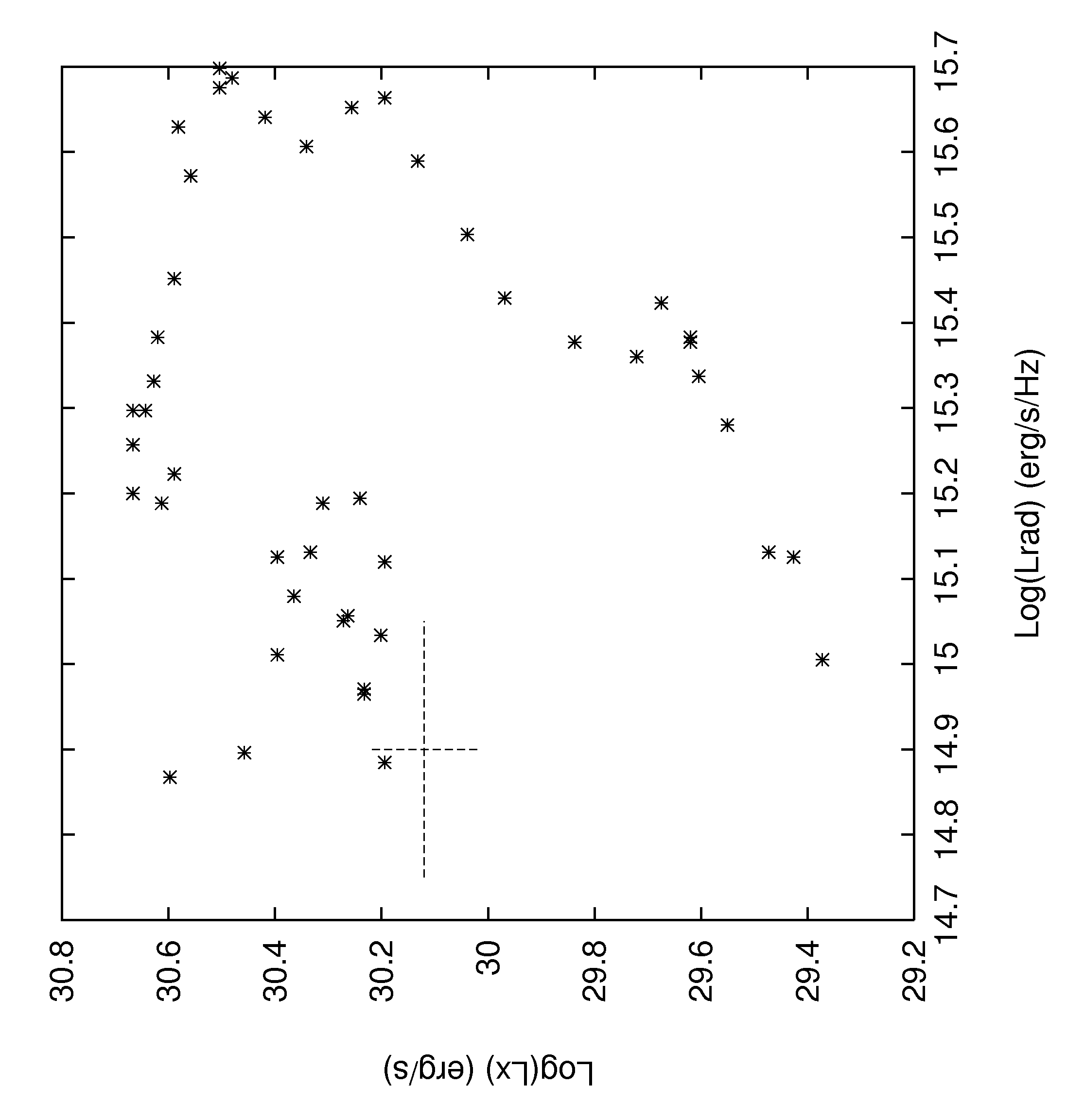}
\caption{ A scanned version of part of Figure~3 from Forbrich et al.\ (2011), covering
observed microwave and X-ray log(flux) combinations for CC Eri flares, is here shown with
a cross at 14.9, 30.12 to represent range of log (flux) values in X-rays and C-band for the flares of AB Dor 
discussed herein.  It would appear that our results for AB Dor are well within the
general correlation discussed by Forbrich et al., and while comparable to those of CC Eri 
are somewhat to the faint side of the distribution for this class of object.} 
\label{AB_Dor_xradcor}
\end{figure}

Apart from these intercorrelations of the microwave data-sets themselves, there is interest in the correlation that such data may have with contemporaneous observations at higher photon energies (Forbrich et al., 2011).  Such correlations can illuminate physical understanding of flare energetics.  Figure~3 in Forbrich et al.\ (2011) shows detailed plots of the microwave variations against contemporaneous X-ray observations for 5 well-known active stars, including the physically 
comparable object CC Eri, mentioned above.  In fact, if we use the measured flare fluxes: in the X-ray region from Section 2.1, and in the C-band from Table 2, and convert to the same (cgs) luminosity scales used in Forbrich et al.'s Figure~3, we will find representative points for these AB Dor flares (X-ray: 1.1 - 1.6$\times$10$^{30}$; 
C-band: 5 - 10$\times$10$^{14}$ Hz$^{-1}$) reasonably typical among the accumulation 
of points for CC Eri towards the lower left of the diagram  (see Figure~\ref{AB_Dor_xradcor}).

The existence of such correlations between X-ray fluxes and microwave emission is taken, in a general way, to support the view that the radio emission originates from the tail of the Maxwellian electron distribution of a very hot plasma through the gyrosynchrotron process; the same, flare-generating, electron population being involved in both emissions.  Although this model has the appeal of physical unification, detailed matching of the quantities involved leads to
complications associated with the structure, properties and geometry
of the sources.  These are not the same for all events, as discussed below.

\section{Discussion}

\begin{figure*}
\hspace{0cm}
\includegraphics[width=12cm,angle=-90.]{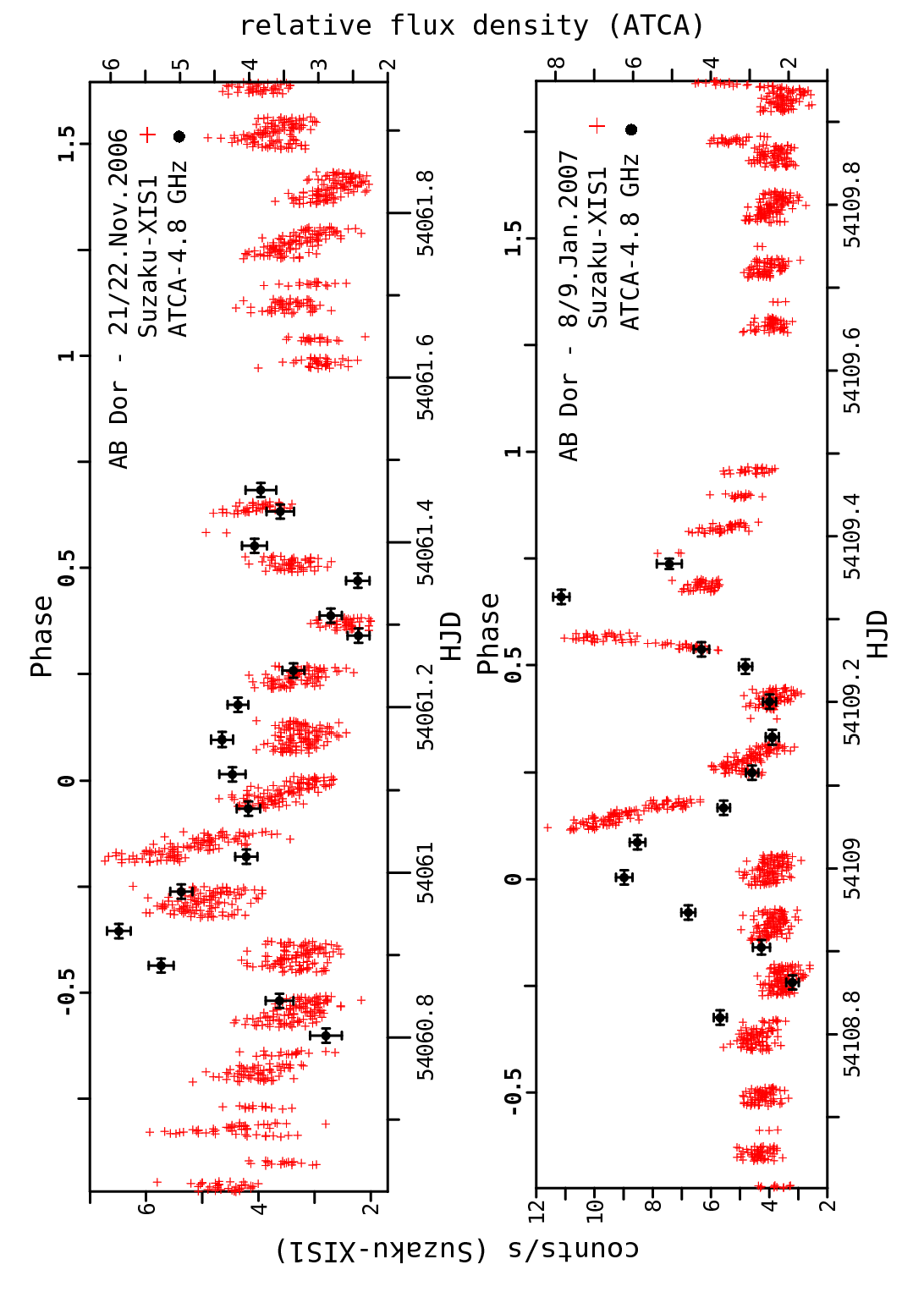}
\caption{Superposition of the microwave (4.8 GHz) and X-ray light curves for AB Dor.
The vertical bars on the microwave points indicate measurement uncertainties, while
the horizontal ones show the duration of the corresponding integrations.}
\label{AB_Dor_xfigt}
\end{figure*}

Budding et al.\ (2009) reviewed in some detail photometry and spectroscopy of AB Dor that formed a subset of the present multiwavelength study.  
The broadband light curves were dominated by one outstanding macula, while the spectroscopic features suggested several chromospheric activity sites, one strong concentration of which, located near phase zero, could be associated with the main macula.  These findings led to the surmise of a large bipolar surface structure that was linked to the site of contemporaneous multiwavelength activity.  Our subsequent reduction and analysis of the Suzaku data are compromised, to some extent, by the lack of a clear resolution between the two
sources.  This certainly holds for the quiescent emission levels, where previous long duration
observations by the Chandra satellite point to the source AB Dor A predominating by a factor of
typically $\sim$6. Moreover, our present X-ray data
show clear links between the major X-ray and microwave flares, which support the idea of one large magnetodynamic activity generator that may be given an approximate position on the surface of AB Dor during late 2006 and early 2007.

The X-ray data, newly presented in this article,
may have a special role in placing the sequence of 
events in flare energization.  In this connection,
X-ray spectral indices for the Suzaku data for  Nov 2006 and Jan 2007  were measured and their behaviour shown in Figure~4.
The data cover quiescent and flaring phases. Of special interest are the near peak regions, particularly the second flare of the Jan 2007 data that shows an initial impulse and subsequent second peak occurring after the 
intermediate microwave peak (see Figure~\ref{AB_Dor_xfigt}). 

The higher-energy quiescent fluxes (not reliably measurable in the highest frequency range for  quiescent GTIs) were consistent with the full X-ray spectra shown in Figure~2, 
and the formula $F_\nu \sim A.\nu^{\delta}$.  The spectral index $\delta$ = --3 is often adopted in theoretical 
treatments  and our  X-ray spectral indices are broadly in agreement with this. 
 Dulk (1985) shows the variation of absorption and emission coefficients for microwave radiation to be quite sensitive
to the value of $\delta$ for gyrosynchrotron emission.  The effective temperature has a moderate dependence, while the
degree of circular polarization is less affected.  It would clearly be useful to take into account observational evidence  
on $\delta$ in more detailed modelling of the radio emission.

An interesting point arising from the juxtaposition of the X-ray and microwave
flux variations (Fig.~\ref{AB_Dor_xfigt})
is whether the initial sharp impulse on
the second flare of Jan 8 (GTI 12) corresponds to
a high-energy trigger preceding the microwave maxima.
This behaviour has been seen in previous similar event sequences 
(Budding et al., 2006).
If we compare the spectral indices for GTIs 9, 12 and 13 (Jan 8/9) there is not
much difference between those of 9 and 12 ---
i.e.\ the initial impulse of the second (smaller) flare is no harder
than the decline phase of the first flare. That points to
 a much more powerful initial impulse to that first flare 
 that was not detected during the observations.
All these GTIs do, however, have significantly harder
spectral indices than the quiescent spectrum.
The index hardens to around  0.5 greater than during quiescence, maximizing, 
in these data sets, at the 
earliest stage of the second flare.
On the other hand, that second flare of Jan 8 is
definitely softer after the intervening microwave peak.
The X-ray spectral  index declines by about 30\% in the second X-ray maximum,
which can be associated with near-footpoint bremsstrahlung radiation
(Bastian et al., 1998).
 The gentler downward slope of the X-rays 
in the post-microwave maxima phases of this flare concurs with the Neupert effect here.
A reasonably self-consistent overall picture for the observed flare sequence
can then be presented, on the basis that there was a first
very large but unseen impulse to the first flare, while for the second
(weaker) flare we observe the whole sequence: initial impulse,
gyrosynchrotron cooling and then bremsstrahlung from precipitated electrons.

This picture can be placed alongside the interpretation of 
optical and radio data given by Budding et al.\ (2009), where it was 
found reasonable to associate the radio observations of 2006 November and 2007 January to one large and predominating visual macula of around 14-15 deg angular radius, longitude 5-10 deg and latitude around 50-60 deg.  A large
chromospheric facular region could also be linked to this macula.
This active region comes into full view near phase zero, when the first of the Jan 2007
flares occurs.  The second major flare of the Jan 2007 observations (as well as the Nov 2006 one), however, occurs around phases 0.6-0.7, which is about the location of `facula b', the second major chromospheric feature 
detected from the H$_{\alpha}$ emission line structure by Budding et al.\ (2009).
  Although about the same, or slightly  weaker, in X-rays (the first flare's 
  X-ray start looks to have been missed), this flare has a stronger microwave 
peak.  Although this flare may then be of intrinsically lower energy
than the first one, a reasonable inference is that the source geometry happens 
to give us a better view, including its initial 
impulsive phase.

It would be unwise to put too much weight on this scenario from just the limited amount of data presented here -- but our interpretation
is worth retaining for further testing against more data that should be
expected in future.   
 
\section*{Acknowledgments}
We thank the time allocation committee of the Australia Telescope National Facility for two generous allocations on the ATCA to support this multiwavelength campaign.  The original proposal and arrangements for acquiring time on the Suzaku satellite actually came from Dr Mark Audard, of Geneva; to whom we express
our gratitude for his early inputs and suggestions.  Dr Grigor Tsarevsky of the Sternberg Institute, Moscow also contributed greatly with advice and observational assistance. 
We thank also the staff of the Armagh Observatory, who helped N.E.\
progress this work, particularly Dr G.\ Ramsay in relation to the
processing of the Suzaku data and subsequent comments on
its discussion.  N.E.'s study leave in Armagh was 
supported by the EU Erasmus Programme.
Detailed comments from a reviewer have led to 
a much more substantiated presentation.


\end{document}